\documentclass[onecolumn]{llncs}

\usepackage[
  pdfauthor={Sam Tobin-Hochstadt},
  pdftitle={Extensible Pattern Matching in an Extensible Language},
  pdfsubject={Computer Science},
  pdfkeywords={Pattern Matching, Extensible Languages}]
{hyperref}

\usepackage{graphicx}
\usepackage{hyperref}

\usepackage{stabular}
\usepackage{relsize}
\usepackage{wasysym}
\usepackage{skull}
\usepackage{textcomp}
\usepackage[htt]{hyphenat}
\usepackage[usenames,dvipsnames]{color}
\hypersetup{bookmarks=true,bookmarksopen=true,bookmarksnumbered=true}

\newcommand{\sectionNewpage}{}

\newcommand{\preDoc}{}
\newcommand{\postDoc}{}

\newcommand{\ChapRefUC}[2]{\SecRefUC{#1}{#2}}
\newcommand{\SecRefUC}[2]{Section~#1}

\newcommand{\Scribtexttt}[1]{{\texttt{#1}}}

\newcommand{\incolorbox}[2]{{\fboxrule=0pt\fboxsep=0pt\colorbox{#1}{#2}}}

\newcommand{\badlink}[1]{#1}

\newcommand{\imageleft}[1]{} 

\newcommand{\planetName}[1]{PLane\hspace{-0.1ex}T}

\newenvironment{bigtabular}{\begin{stabular}}{\end{stabular}}

\newlength{\stabLeft}
\newcommand{\bigtableleftpad}{\hspace{\stabLeft}}
\newcommand{\atItemizeStart}[0]{\addtolength{\stabLeft}{\labelsep}
                                \addtolength{\stabLeft}{\labelwidth}}

\newenvironment{SingleColumn}{\begin{list}{}{\topsep=0pt\partopsep=0pt%
\listparindent=0pt\itemindent=0pt\labelwidth=0pt\leftmargin=0pt\rightmargin=0pt%
\itemsep=0pt\parsep=0pt}\item}{\end{list}}

\newenvironment{Subflow}{\begin{list}{}{\topsep=0pt\partopsep=0pt%
\listparindent=0pt\itemindent=0pt\labelwidth=0pt\leftmargin=0pt\rightmargin=0pt%
\itemsep=0pt}\item}{\end{list}}

\newcommand{\SCodePreSkip}{\vskip\abovedisplayskip}
\newcommand{\SCodePostSkip}{\vskip\belowdisplayskip}
\newenvironment{SCodeFlow}{\SCodePreSkip\begin{list}{}{\topsep=0pt\partopsep=0pt%
\listparindent=0pt\itemindent=0pt\labelwidth=0pt\leftmargin=2ex\rightmargin=0pt%
\itemsep=0pt\parsep=0pt}\item}{\end{list}\SCodePostSkip}

\newenvironment{refcolumn}{}{}

\newcommand{\titleAndEmptyVersionAndEmptyAuthors}[3]{\title{#1}\maketitle}

\newcommand{\SAuthorSep}[1]{\qquad}

\newcommand{\SColorize}[2]{\color{#1}{#2}}

\newcommand{\inColor}[2]{{\Scribtexttt{\SColorize{#1}{#2}}}}
\definecolor{PaleBlue}{rgb}{0.90,0.90,1.0}
\definecolor{LightGray}{rgb}{0.90,0.90,0.90}
\definecolor{CommentColor}{rgb}{0.76,0.45,0.12}
\definecolor{ParenColor}{rgb}{0.52,0.24,0.14}
\definecolor{IdentifierColor}{rgb}{0.15,0.15,0.50}
\definecolor{ResultColor}{rgb}{0.0,0.0,0.69}
\definecolor{ValueColor}{rgb}{0.13,0.55,0.13}
\definecolor{OutputColor}{rgb}{0.59,0.00,0.59}

\newcommand{\RktPlain}[1]{\inColor{black}{#1}}
\newcommand{\RktKw}[1]{{\SColorize{black}{\Scribtexttt{#1}}}} 
\newcommand{\RktStxLink}[1]{\RktKw{#1}}
\newcommand{\RktCmt}[1]{\inColor{CommentColor}{#1}}
\newcommand{\RktPn}[1]{\inColor{ParenColor}{#1}}
\newcommand{\RktInBG}[1]{\inColor{ParenColor}{#1}}
\newcommand{\RktSym}[1]{\inColor{IdentifierColor}{#1}}
\newcommand{\RktVal}[1]{\inColor{ValueColor}{#1}}
\newcommand{\RktValLink}[1]{\inColor{blue}{#1}}

\newcommand{\RktRes}[1]{\inColor{ResultColor}{#1}}
\newcommand{\RktOut}[1]{\inColor{OutputColor}{#1}}

\newcommand{\RktRdr}[1]{\inColor{black}{#1}}

\newcommand{\RktErrCol}[1]{\inColor{red}{#1}}
\newcommand{\RktErr}[1]{{\RktErrCol{\textrm{\textit{#1}}}}}

\newcommand{\RktIn}[1]{\incolorbox{LightGray}{\RktInBG{#1}}}

\newenvironment{RktBlk}{}{}

\usepackage{ccaption}

\newcommand{\Legend}[1]{~

                        \hrule width \hsize height .33pt
                        \vspace{4pt}
                        \legend{#1}}

\newlength{\FigOrigskip}
\FigOrigskip=\parskip

\newenvironment{CenterfigureMulti}{\begin{figure*}[t!p]\centering}{\end{figure*}}

\newenvironment{Centerfigure}{\begin{figure}[t!p]\centering}{\end{figure}}

\newenvironment{FigureInside}{\begin{list}{}{\leftmargin=0pt\topsep=0pt\parsep=\FigOrigskip\partopsep=0pt}\item}{\end{list}}

\author{Sam Tobin-Hochstadt}
\institute{PLT @ Northeastern University\\
           {\tt samth@ccs.neu.edu}}

\renewcommand{\sectionNewpage}{}

\newcommand\overmath[1]{\ensuremath{\overline{#1}}}

\begin{document}
\preDoc
\titleAndEmptyVersionAndEmptyAuthors{Extensible Pattern Matching in an Extensible Language}{}{}
\label{t:x28part_x22Extensiblex5fPatternx5fMatchingx5finx5fanx5fExtensiblex5fLanguagex22x29}

\begin{abstract}Pattern matching is a widely used technique in
functional languages, especially those in the ML and
Haskell traditions, where it is at the core of the semantics.
  In languages in the Lisp tradition, in contrast, pattern
matching it typically provided by
libraries built with macros.  We
present \RktSym{\RktStxLink{match}}, a sophisticated pattern matcher for Racket, implemented as
language extension.
using macros.  The system supports novel and widely{-}useful pattern{-}matching
forms, and is itself extensible. The extensibility of
\RktSym{\RktStxLink{match}} is implemented via a general technique for creating
extensible language extensions.\end{abstract}

\sectionNewpage

\section[Extending Pattern Matching]{Extending Pattern Matching}\label{t:x28part_x22introx22x29}

The following Racket\footnote{Racket is the new name of PLT Scheme.}~\cite{plttr1}
 program finds the magnitude of a complex number,
represented in either Cartesian or polar form as a 3{-}element list,
 using the first element as a type tag:

\begin{SCodeFlow}\begin{RktBlk}\begin{SingleColumn}\RktPn{(}\RktSym{\RktStxLink{define}}\mbox{\hphantom{\Scribtexttt{x}}}\RktPn{(}\RktSym{\RktValLink{magnitude}}\mbox{\hphantom{\Scribtexttt{x}}}\RktSym{n}\RktPn{)}

\mbox{\hphantom{\Scribtexttt{xx}}}\RktPn{(}\RktSym{\RktStxLink{cond}}\mbox{\hphantom{\Scribtexttt{x}}}\RktPn{[}\RktPn{(}\RktSym{\RktValLink{eq{\hbox{\texttt{?}}}}}\mbox{\hphantom{\Scribtexttt{x}}}\RktPn{(}\RktSym{\RktValLink{first}}\mbox{\hphantom{\Scribtexttt{x}}}\RktSym{n}\RktPn{)}\mbox{\hphantom{\Scribtexttt{x}}}\RktVal{{'}}\RktVal{cart}\RktPn{)}

\mbox{\hphantom{\Scribtexttt{xxxxxxxxx}}}\RktPn{(}\RktSym{\RktValLink{sqrt}}\mbox{\hphantom{\Scribtexttt{x}}}\RktPn{(}\RktSym{\RktValLink{+}}\mbox{\hphantom{\Scribtexttt{x}}}\RktPn{(}\RktSym{\RktValLink{sqr}}\mbox{\hphantom{\Scribtexttt{x}}}\RktPn{(}\RktSym{\RktValLink{second}}\mbox{\hphantom{\Scribtexttt{x}}}\RktSym{n}\RktPn{)}\RktPn{)}\mbox{\hphantom{\Scribtexttt{x}}}\RktPn{(}\RktSym{\RktValLink{sqr}}\mbox{\hphantom{\Scribtexttt{x}}}\RktPn{(}\RktSym{\RktValLink{third}}\mbox{\hphantom{\Scribtexttt{x}}}\RktSym{n}\RktPn{)}\RktPn{)}\RktPn{)}\RktPn{)}\RktPn{]}

\mbox{\hphantom{\Scribtexttt{xxxxxxxx}}}\RktPn{[}\RktPn{(}\RktSym{\RktValLink{eq{\hbox{\texttt{?}}}}}\mbox{\hphantom{\Scribtexttt{x}}}\RktPn{(}\RktSym{\RktValLink{first}}\mbox{\hphantom{\Scribtexttt{x}}}\RktSym{n}\RktPn{)}\mbox{\hphantom{\Scribtexttt{x}}}\RktVal{{'}}\RktVal{polar}\RktPn{)}

\mbox{\hphantom{\Scribtexttt{xxxxxxxxx}}}\RktPn{(}\RktSym{\RktValLink{second}}\mbox{\hphantom{\Scribtexttt{x}}}\RktSym{n}\RktPn{)}\RktPn{]}\RktPn{)}\RktPn{)}\end{SingleColumn}\end{RktBlk}\end{SCodeFlow}

\noindent While this program accomplishes the desired purpose, it{'}s far from
obviously correct, and commits the program to the list{-}based representation.
  Additionally, it unnecessarily repeats  accesses to the list structure
making up the representation.  Finally, if the input is
\RktVal{{'}}\RktVal{(}\RktVal{cart}\Scribtexttt{ }\RktVal{7}\RktVal{)}, it produces a hard{-}to{-}decipher error from the
\RktSym{\RktValLink{third}} function.

In contrast, the same program written using pattern matching is far
simpler:

\begin{SCodeFlow}\begin{RktBlk}\begin{SingleColumn}\RktPn{(}\RktSym{\RktStxLink{define}}\mbox{\hphantom{\Scribtexttt{x}}}\RktPn{(}\RktSym{\RktValLink{magnitude}}\mbox{\hphantom{\Scribtexttt{x}}}\RktSym{n}\RktPn{)}

\mbox{\hphantom{\Scribtexttt{xx}}}\RktPn{(}\RktSym{\RktStxLink{match}}\mbox{\hphantom{\Scribtexttt{x}}}\RktSym{n}

\mbox{\hphantom{\Scribtexttt{xxxx}}}\RktPn{[}\RktPn{(}\RktSym{\RktValLink{list}}\mbox{\hphantom{\Scribtexttt{x}}}\RktVal{{'}}\RktVal{cart}\mbox{\hphantom{\Scribtexttt{x}}}\RktSym{x}\mbox{\hphantom{\Scribtexttt{x}}}\RktSym{y}\RktPn{)}\mbox{\hphantom{\Scribtexttt{x}}}\RktPn{(}\RktSym{\RktValLink{sqrt}}\mbox{\hphantom{\Scribtexttt{x}}}\RktPn{(}\RktSym{\RktValLink{+}}\mbox{\hphantom{\Scribtexttt{x}}}\RktPn{(}\RktSym{\RktValLink{sqr}}\mbox{\hphantom{\Scribtexttt{x}}}\RktSym{x}\RktPn{)}\mbox{\hphantom{\Scribtexttt{x}}}\RktPn{(}\RktSym{\RktValLink{sqr}}\mbox{\hphantom{\Scribtexttt{x}}}\RktSym{y}\RktPn{)}\RktPn{)}\RktPn{)}\RktPn{]}

\mbox{\hphantom{\Scribtexttt{xxxx}}}\RktPn{[}\RktPn{(}\RktSym{\RktValLink{list}}\mbox{\hphantom{\Scribtexttt{x}}}\RktVal{{'}}\RktVal{polar}\mbox{\hphantom{\Scribtexttt{x}}}\RktSym{r}\mbox{\hphantom{\Scribtexttt{x}}}\RktSym{theta}\RktPn{)}\mbox{\hphantom{\Scribtexttt{x}}}\RktSym{r}\RktPn{]}\RktPn{)}\RktPn{)}\end{SingleColumn}\end{RktBlk}\end{SCodeFlow}

\noindent The new program is shorter, more perspicuous, does not repeat computation,
and produces better error messages.  For this reason, pattern matching
has become a ubiquitous tool in functional programming, especially for
languages in the Haskell and ML families.  Unfortunately, pattern
matching is less ubiquitous in functional languages in the Lisp tradition,
such as Common Lisp, Scheme, and Racket.  This is unfortunate, since
as we demonstrate in the remainder of the paper,
 not only are the same benefits available as in Haskell or ML, but the
extensibility provided by languages such as Racket leads naturally to
expressive and extensible pattern matchers.

\subsection[More Expressive Patterns]{More Expressive Patterns}\label{t:x28part_x22Morex5fExpressivex5fPatternsx22x29}

The function can also be easily converted to arbitrary{-}dimensional
coordinates, using the \RktSym{\RktStxLink{{\hbox{\texttt{.}}}{\hbox{\texttt{.}}}{\hbox{\texttt{.}}}}} notation for specifying an arbitrary{-}length
list:

\noindent \begin{SCodeFlow}\begin{RktBlk}\begin{SingleColumn}\RktPn{(}\RktSym{\RktStxLink{define}}\mbox{\hphantom{\Scribtexttt{x}}}\RktPn{(}\RktSym{\RktValLink{magnitude}}\mbox{\hphantom{\Scribtexttt{x}}}\RktSym{n}\RktPn{)}

\mbox{\hphantom{\Scribtexttt{xx}}}\RktPn{(}\RktSym{\RktStxLink{match}}\mbox{\hphantom{\Scribtexttt{x}}}\RktSym{n}

\mbox{\hphantom{\Scribtexttt{xxxx}}}\RktPn{[}\RktPn{(}\RktSym{\RktValLink{list}}\mbox{\hphantom{\Scribtexttt{x}}}\RktVal{{'}}\RktVal{cart}\mbox{\hphantom{\Scribtexttt{x}}}\RktSym{xs}\mbox{\hphantom{\Scribtexttt{x}}}\RktSym{\RktStxLink{{\hbox{\texttt{.}}}{\hbox{\texttt{.}}}{\hbox{\texttt{.}}}}}\RktPn{)}\mbox{\hphantom{\Scribtexttt{x}}}\RktPn{(}\RktSym{\RktValLink{sqrt}}\mbox{\hphantom{\Scribtexttt{x}}}\RktPn{(}\RktSym{\RktValLink{apply}}\mbox{\hphantom{\Scribtexttt{x}}}\RktSym{\RktValLink{+}}\mbox{\hphantom{\Scribtexttt{x}}}\RktPn{(}\RktSym{\RktValLink{map}}\mbox{\hphantom{\Scribtexttt{x}}}\RktSym{\RktValLink{sqr}}\mbox{\hphantom{\Scribtexttt{x}}}\RktSym{xs}\RktPn{)}\RktPn{)}\RktPn{)}\RktPn{]}

\mbox{\hphantom{\Scribtexttt{xxxx}}}\RktPn{[}\RktPn{(}\RktSym{\RktValLink{list}}\mbox{\hphantom{\Scribtexttt{x}}}\RktVal{{'}}\RktVal{polar}\mbox{\hphantom{\Scribtexttt{x}}}\RktSym{r}\mbox{\hphantom{\Scribtexttt{x}}}\RktSym{theta}\mbox{\hphantom{\Scribtexttt{x}}}\RktSym{\RktStxLink{{\hbox{\texttt{.}}}{\hbox{\texttt{.}}}{\hbox{\texttt{.}}}}}\RktPn{)}\mbox{\hphantom{\Scribtexttt{x}}}\RktSym{r}\RktPn{]}\RktPn{)}\RktPn{)}\end{SingleColumn}\end{RktBlk}\end{SCodeFlow}

\noindent Racket is untyped, so we can add argument checking in the pattern
match to catch errors early. Here we use the \RktKw{{\hbox{\texttt{?}}}} pattern, which tests
the value under consideration against the supplied predicate:

\noindent \begin{SCodeFlow}\begin{RktBlk}\begin{SingleColumn}\RktPn{(}\RktSym{\RktStxLink{define}}\mbox{\hphantom{\Scribtexttt{x}}}\RktPn{(}\RktSym{\RktValLink{magnitude}}\mbox{\hphantom{\Scribtexttt{x}}}\RktSym{n}\RktPn{)}

\mbox{\hphantom{\Scribtexttt{xx}}}\RktPn{(}\RktSym{\RktStxLink{match}}\mbox{\hphantom{\Scribtexttt{x}}}\RktSym{n}

\mbox{\hphantom{\Scribtexttt{xxxx}}}\RktPn{[}\RktPn{(}\RktSym{\RktValLink{list}}\mbox{\hphantom{\Scribtexttt{x}}}\RktVal{{'}}\RktVal{cart}\mbox{\hphantom{\Scribtexttt{x}}}\RktPn{(}\RktKw{{\hbox{\texttt{?}}}}\mbox{\hphantom{\Scribtexttt{x}}}\RktSym{\RktValLink{real{\hbox{\texttt{?}}}}}\mbox{\hphantom{\Scribtexttt{x}}}\RktSym{xs}\RktPn{)}\mbox{\hphantom{\Scribtexttt{x}}}\RktSym{\RktStxLink{{\hbox{\texttt{.}}}{\hbox{\texttt{.}}}{\hbox{\texttt{.}}}}}\RktPn{)}\mbox{\hphantom{\Scribtexttt{x}}}\RktPn{(}\RktSym{\RktValLink{sqrt}}\mbox{\hphantom{\Scribtexttt{x}}}\RktPn{(}\RktSym{\RktValLink{apply}}\mbox{\hphantom{\Scribtexttt{x}}}\RktSym{\RktValLink{+}}\mbox{\hphantom{\Scribtexttt{x}}}\RktPn{(}\RktSym{\RktValLink{map}}\mbox{\hphantom{\Scribtexttt{x}}}\RktSym{\RktValLink{sqr}}\mbox{\hphantom{\Scribtexttt{x}}}\RktSym{xs}\RktPn{)}\RktPn{)}\RktPn{)}\RktPn{]}

\mbox{\hphantom{\Scribtexttt{xxxx}}}\RktPn{[}\RktPn{(}\RktSym{\RktValLink{list}}\mbox{\hphantom{\Scribtexttt{x}}}\RktVal{{'}}\RktVal{polar}\mbox{\hphantom{\Scribtexttt{x}}}\RktPn{(}\RktKw{{\hbox{\texttt{?}}}}\mbox{\hphantom{\Scribtexttt{x}}}\RktSym{\RktValLink{real{\hbox{\texttt{?}}}}}\mbox{\hphantom{\Scribtexttt{x}}}\RktSym{r}\RktPn{)}\mbox{\hphantom{\Scribtexttt{x}}}\RktPn{(}\RktKw{{\hbox{\texttt{?}}}}\mbox{\hphantom{\Scribtexttt{x}}}\RktSym{\RktValLink{real{\hbox{\texttt{?}}}}}\mbox{\hphantom{\Scribtexttt{x}}}\RktSym{theta}\RktPn{)}\mbox{\hphantom{\Scribtexttt{x}}}\RktSym{\RktStxLink{{\hbox{\texttt{.}}}{\hbox{\texttt{.}}}{\hbox{\texttt{.}}}}}\RktPn{)}\mbox{\hphantom{\Scribtexttt{x}}}\RktSym{r}\RktPn{]}\RktPn{)}\RktPn{)}\end{SingleColumn}\end{RktBlk}\end{SCodeFlow}

\noindent This implementation is more robust than our original function, but it
approaches it in complexity, and still commits us to a list{-}based
representation of coordinates.

\subsection[Custom Patterns]{Custom Patterns}\label{t:x28part_x22Customx5fPatternsx22x29}

By
switching to custom, user{-}defined pattern matching forms, we can
  simplify our patterns and the
representation choice can be abstracted away:

\noindent \begin{SCodeFlow}\begin{RktBlk}\begin{SingleColumn}\RktPn{(}\RktSym{\RktStxLink{define}}\mbox{\hphantom{\Scribtexttt{x}}}\RktPn{(}\RktSym{\RktValLink{magnitude}}\mbox{\hphantom{\Scribtexttt{x}}}\RktSym{n}\RktPn{)}

\mbox{\hphantom{\Scribtexttt{xx}}}\RktPn{(}\RktSym{\RktStxLink{match}}\mbox{\hphantom{\Scribtexttt{x}}}\RktSym{n}

\mbox{\hphantom{\Scribtexttt{xxxx}}}\RktPn{[}\RktPn{(}\RktSym{\badlink{\RktValLink{cart}}}\mbox{\hphantom{\Scribtexttt{x}}}\RktSym{xs}\mbox{\hphantom{\Scribtexttt{x}}}\RktSym{\RktStxLink{{\hbox{\texttt{.}}}{\hbox{\texttt{.}}}{\hbox{\texttt{.}}}}}\RktPn{)}

\mbox{\hphantom{\Scribtexttt{xxxxx}}}\RktPn{(}\RktSym{\RktValLink{sqrt}}\mbox{\hphantom{\Scribtexttt{x}}}\RktPn{(}\RktSym{\RktValLink{apply}}\mbox{\hphantom{\Scribtexttt{x}}}\RktSym{\RktValLink{+}}\mbox{\hphantom{\Scribtexttt{x}}}\RktPn{(}\RktSym{\RktValLink{map}}\mbox{\hphantom{\Scribtexttt{x}}}\RktSym{\RktValLink{sqr}}\mbox{\hphantom{\Scribtexttt{x}}}\RktSym{xs}\RktPn{)}\RktPn{)}\RktPn{)}\RktPn{]}

\mbox{\hphantom{\Scribtexttt{xxxx}}}\RktPn{[}\RktPn{(}\RktSym{\badlink{\RktValLink{polar}}}\mbox{\hphantom{\Scribtexttt{x}}}\RktSym{r}\mbox{\hphantom{\Scribtexttt{x}}}\RktSym{theta}\mbox{\hphantom{\Scribtexttt{x}}}\RktSym{\RktStxLink{{\hbox{\texttt{.}}}{\hbox{\texttt{.}}}{\hbox{\texttt{.}}}}}\RktPn{)}\mbox{\hphantom{\Scribtexttt{x}}}\RktSym{r}\RktPn{]}\RktPn{)}\RktPn{)}\end{SingleColumn}\end{RktBlk}\end{SCodeFlow}

Our custom pattern matching form can use other features of Racket{'}s
pattern matcher to perform arbitrary computation,
allowing us to simplify the function further by transparently
converting Cartesian to polar coordinates when necessary:

\noindent \begin{SCodeFlow}\begin{RktBlk}\begin{SingleColumn}\RktPn{(}\RktSym{\RktStxLink{define}}\mbox{\hphantom{\Scribtexttt{x}}}\RktPn{(}\RktSym{\RktValLink{magnitude}}\mbox{\hphantom{\Scribtexttt{x}}}\RktSym{n}\RktPn{)}

\mbox{\hphantom{\Scribtexttt{xx}}}\RktPn{(}\RktSym{\RktStxLink{match}}\mbox{\hphantom{\Scribtexttt{x}}}\RktSym{n}

\mbox{\hphantom{\Scribtexttt{xxxx}}}\RktPn{[}\RktPn{(}\RktSym{\badlink{\RktValLink{polar}}}\mbox{\hphantom{\Scribtexttt{x}}}\RktSym{r}\mbox{\hphantom{\Scribtexttt{x}}}\RktSym{theta}\mbox{\hphantom{\Scribtexttt{x}}}\RktSym{\RktStxLink{{\hbox{\texttt{.}}}{\hbox{\texttt{.}}}{\hbox{\texttt{.}}}}}\RktPn{)}\mbox{\hphantom{\Scribtexttt{x}}}\RktSym{r}\RktPn{]}\RktPn{)}\RktPn{)}\end{SingleColumn}\end{RktBlk}\end{SCodeFlow}

\noindent We now have an implementation which avoids commitment to
representation and handles arbitrary dimensions as well as argument
checking.
  Without extensible pattern
matching, this might require interface changes or a large number of
helper functions.  Instead, we have developed reusable abstractions
which support the clear definition of exactly what we mean.

In the remainder of the paper, we describe the implementation of all
of these examples, focusing on user{-}extensibility.
We begin with a tour of pattern matching in Racket, touching
briefly on the adaptation of standard techniques from ML{-}style
matching~\cite{maranget} and their implementation via macros~\cite{dhb:sc,ctf:macros}.
Then we describe the implementation of sequence patterns{---}seen above
with the use of \RktSym{\RktStxLink{{\hbox{\texttt{.}}}{\hbox{\texttt{.}}}{\hbox{\texttt{.}}}}}{---}and other pattern forms not found in conventional
pattern{-}matching systems.  Third, we describe how to make patterns
user{-}extensible by exploiting the flexibility of
Racket{'}s macro system.  Finally, we discuss related work, including a
history of pattern matching in Scheme and Racket.

\sectionNewpage

\section[Pattern Matching in Racket]{Pattern Matching in Racket}\label{t:x28part_x22basicx22x29}

\begin{Centerfigure}\begin{FigureInside}\begin{bigtabular}{@{\bigtableleftpad}l@{}l@{}l@{}l@{}}
 &
\relax{\quad\ $|$ } &
\RktSym{x} &
\begin{minipage}{0.25\linewidth}
\begin{Subflow}\mbox{\hphantom{\Scribtexttt{xx}}}variables\end{Subflow} \end{minipage}
 \\
\RktSym{pat} &
\relax{\quad   $::=$  } &
\RktPn{(}\RktKw{{\hbox{\texttt{?}}}}\Scribtexttt{ }\RktSym{\RktStxLink{expr}}\RktPn{)} &
\begin{minipage}{0.25\linewidth}
\begin{Subflow}\mbox{\hphantom{\Scribtexttt{xx}}}predicates\end{Subflow} \end{minipage}
 \\
 &
\relax{\quad\ $|$ } &
\RktPn{(}\RktSym{\RktStxLink{and}}\Scribtexttt{ }\overmath{\RktSym{pat}}\RktPn{)} &
\begin{minipage}{0.25\linewidth}
\begin{Subflow}\mbox{\hphantom{\Scribtexttt{xx}}}conjunction\end{Subflow} \end{minipage}
 \\
 &
\relax{\quad\ $|$ } &
\RktVal{{'}}\RktSym{val} &
\begin{minipage}{0.25\linewidth}
\begin{Subflow}\mbox{\hphantom{\Scribtexttt{xx}}}literal data\end{Subflow} \end{minipage}
 \\
 &
\relax{\quad\ $|$ } &
\RktPn{(}\RktKw{cons}\Scribtexttt{ }\RktSym{pat}\Scribtexttt{ }\RktSym{pat}\RktPn{)} &
\begin{minipage}{0.25\linewidth}
\begin{Subflow}\mbox{\hphantom{\Scribtexttt{xx}}}pairs\end{Subflow} \end{minipage}
 \\
 &
\relax{\quad\ $|$ } &
\RktPn{(}\RktKw{list}\Scribtexttt{ }\overmath{\RktSym{pat}}\RktPn{)} &
\begin{minipage}{0.25\linewidth}
\begin{Subflow}\mbox{\hphantom{\Scribtexttt{xx}}}fixed{-}length lists\end{Subflow} \end{minipage}
\end{bigtabular}

\Legend{\label{t:x28counter_x28x22figurex22_x22figx3asimplex2dpatx22x29x29}Figure~1: Simple Patterns}\end{FigureInside}\end{Centerfigure}

In this section, we describe both the interface and implementation of
the most basic pattern matching constructs in Racket.  The fundamental
pattern matching form is \RktSym{\RktStxLink{match}}, with the syntax

\begin{SCodeFlow}\RktPn{(}\RktSym{\RktStxLink{match}}\mbox{\hphantom{\Scribtexttt{x}}}\RktSym{\RktStxLink{expr}}\mbox{\hphantom{\Scribtexttt{x}}}\overmath{\RktPn{[}\RktSym{pat}\Scribtexttt{ }\RktSym{\RktStxLink{expr}}\RktPn{]}}\RktPn{)}\end{SCodeFlow}

\noindent The meaning of \RktSym{\RktStxLink{match}} expressions is the traditional first{-}match
semantics.
 The grammar of some simple patterns is given in
Figure~1.  In reality, \RktSym{\RktStxLink{match}} provides
many more forms for all of the basic data types in Racket, including mutable pairs, vectors,
strings, hash tables, and many others; user{-}defined structures are discussed
in \ChapRefUC{3}{Advanced Patterns}.  The semantics are as usual for pattern
matching: variable patterns match any value and bind the variable in the
right{-}hand side to the value matched, literals (written here with \RktInBG{\RktIn{{'}}})
match just themselves, \RktKw{list}{-} and \RktKw{cons}{-}patterns match structurally,
and \RktSym{\RktStxLink{and}}{-}patterns
match if both conjuncts do.  The only non{-}standard pattern is
\RktPn{(}\RktKw{{\hbox{\texttt{?}}}}\Scribtexttt{ }\RktSym{\RktStxLink{expr}}\RktPn{)}. Here, \RktSym{\RktStxLink{expr}} must evaluate to a one argument
function, and the pattern matches if that function produces
\RktSym{\RktValLink{true}} (or any other non{-}false value)
 when applied to the value being considered
 (the \textit{scrutinee}). For example,
the pattern \RktPn{(}\RktSym{\RktStxLink{and}}\Scribtexttt{ }\RktPn{(}\RktKw{{\hbox{\texttt{?}}}}\Scribtexttt{ }\RktSym{\RktValLink{even{\hbox{\texttt{?}}}}}\RktPn{)}\Scribtexttt{ }\RktSym{x}\RktPn{)} matches only even numbers and
binds them to \RktSym{x}.  A \RktSym{\RktStxLink{match}} expression matches the
scrutinee against each pattern in turn, executing the right{-}hand side
corresponding to the first successful match.  If no pattern matches, a
runtime error is signaled.

\subsection[Compilation to Racket]{Compilation to Racket}\label{t:x28part_x22Compilationx5ftox5fRacketx22x29}

The basic compilation model adopted by \RktSym{\RktStxLink{match}} is the
backtracking automata framework introduced by
Augustsson~\cite{a:fpca} and optimized by Le Fessant and
Maranget~\cite{maranget}.  However, unlike these models,
\RktSym{\RktStxLink{match}} is implemented as a macro, and thus produces
plain Racket code directly, rather than exploiting lower{-}level mechanisms
for tag dispatch and conditional branching. Each automata state is represented by
a thunk (0{-}argument procedure) which may be tail{-}called by a later
computation.  Conditional checks simply use \RktSym{\RktStxLink{if}} tests and
type{-}testing predicates such as \RktSym{\RktValLink{pair{\hbox{\texttt{?}}}}}.

To compile the following \RktSym{\RktStxLink{match}} expression:

\begin{SCodeFlow}\begin{RktBlk}\begin{SingleColumn}\RktPn{(}\RktSym{\RktStxLink{match}}\mbox{\hphantom{\Scribtexttt{x}}}\RktSym{e{\char`\_}1}

\mbox{\hphantom{\Scribtexttt{xx}}}\RktPn{[}\RktPn{(}\RktSym{\RktStxLink{and}}\mbox{\hphantom{\Scribtexttt{x}}}\RktPn{(}\RktKw{{\hbox{\texttt{?}}}}\mbox{\hphantom{\Scribtexttt{x}}}\RktSym{\RktValLink{number{\hbox{\texttt{?}}}}}\RktPn{)}\mbox{\hphantom{\Scribtexttt{x}}}\RktSym{x}\RktPn{)}\mbox{\hphantom{\Scribtexttt{x}}}\RktSym{e{\char`\_}2}\RktPn{]}

\mbox{\hphantom{\Scribtexttt{xx}}}\RktPn{[}\RktSym{\RktStxLink{{\char`\_}}}\mbox{\hphantom{\Scribtexttt{x}}}\RktSym{e{\char`\_}3}\RktPn{]}\RktPn{)}\end{SingleColumn}\end{RktBlk}\end{SCodeFlow}

\noindent the following code is generated:

\begin{SCodeFlow}\begin{RktBlk}\begin{SingleColumn}\RktPn{(}\RktSym{\RktStxLink{let}}\mbox{\hphantom{\Scribtexttt{x}}}\RktPn{(}\RktPn{[}\RktSym{tmp}\mbox{\hphantom{\Scribtexttt{x}}}\RktSym{e{\char`\_}1}\RktPn{]}

\mbox{\hphantom{\Scribtexttt{xxxxxx}}}\RktPn{[}\RktSym{f}\mbox{\hphantom{\Scribtexttt{x}}}\RktPn{(}\RktSym{\RktStxLink{$\lambda$}}\mbox{\hphantom{\Scribtexttt{x}}}\RktPn{(}\RktPn{)}\mbox{\hphantom{\Scribtexttt{x}}}\RktSym{e{\char`\_}3}\RktPn{)}\RktPn{]}\RktPn{)}

\mbox{\hphantom{\Scribtexttt{xx}}}\RktPn{(}\RktSym{\RktStxLink{if}}\mbox{\hphantom{\Scribtexttt{x}}}\RktPn{(}\RktSym{\RktValLink{number{\hbox{\texttt{?}}}}}\mbox{\hphantom{\Scribtexttt{x}}}\RktSym{tmp}\RktPn{)}

\mbox{\hphantom{\Scribtexttt{xxxxxx}}}\RktPn{(}\RktSym{\RktStxLink{let}}\mbox{\hphantom{\Scribtexttt{x}}}\RktPn{(}\RktPn{[}\RktSym{x}\mbox{\hphantom{\Scribtexttt{x}}}\RktSym{tmp}\RktPn{]}\RktPn{)}\mbox{\hphantom{\Scribtexttt{x}}}\RktSym{e{\char`\_}2}\RktPn{)}

\mbox{\hphantom{\Scribtexttt{xxxxxx}}}\RktPn{(}\RktSym{f}\RktPn{)}\RktPn{)}\RktPn{)}\end{SingleColumn}\end{RktBlk}\end{SCodeFlow}

\noindent Here, \RktSym{f} is the failure continuation; it is invoked if the
\RktSym{\RktValLink{number{\hbox{\texttt{?}}}}} test fails.  Testing is a simple conditional, and
variable patterns translate directly to \RktSym{\RktStxLink{let}} binding.  In
larger patterns, multiple failure continuations are generated and may
be called in multiple places.

One important detail is added to the compilation process by the
presence of the \RktKw{{\hbox{\texttt{?}}}} pattern.  Since \RktKw{{\hbox{\texttt{?}}}} patterns contain
an \textit{expression} rather than a subpattern, the expression is
evaluated at some point during matching.  This expression should be
only evaluated once, to avoid needless recomputation as well as
duplicate effects, but should not be computed if it is not needed.
Finally, since backtracking automata may test data multiple times, the
function \textit{produced} by the expression may be called 0, 1, or more
times during matching.

\subsection[Static Checking]{Static Checking}\label{t:x28part_x22Staticx5fCheckingx22x29}

One key design choice is already fixed by the simple patterns, in
particular by the presence of \RktKw{{\hbox{\texttt{?}}}}:
disjointness of patterns and completeness of matching are now impossible to check in the general
case, just as in Haskell with pattern guards.  Even in the absence of \RktKw{{\hbox{\texttt{?}}}}, the untyped nature of Racket
means that checking completeness of pattern matching is only possible in the trivial case when
 a catch{-}all
pattern is provided.  In view of these limitations, \RktSym{\RktStxLink{match}}
does not warn the user when it cannot prove the absence of unreachable
patterns or potentially uncovered cases.

In more restricted systems, such as the
\RktSym{cases} form provided by some textbooks~\cite{eopl3,plai},
these problems become tractable, but the presence of expressive
patterns such as \RktKw{{\hbox{\texttt{?}}}}
patterns makes it impossible here.  We are investigating whether Typed
Racket~\cite{thf:popl08} makes more static checking possible.

\subsection[Implementing Syntactic Extensions]{Implementing Syntactic Extensions}\label{t:x28part_x22Implementingx5fSyntacticx5fExtensionsx22x29}

The basic structure of the \RktSym{\RktStxLink{match}} implementation  is common to many pattern
matching compilers.  The distinctive aspect of the implementation of
Racket{'}s pattern matcher is the use of syntactic extension in the form
of macros to implement the  \RktSym{\RktStxLink{match}} compiler.  We briefly review
Racket{'}s syntactic extension mechanisms and describe how they are used
for implementing \RktSym{\RktStxLink{match}}.

\subsubsection[Defining Syntactic Extensions]{Defining Syntactic Extensions}\label{t:x28part_x22Definingx5fSyntacticx5fExtensionsx22x29}

The fundamental extension form is \RktSym{\RktStxLink{define{-}syntax}}, which binds
values to names in the \textit{syntactic} environment, i.e., the
environment used at compilation time.  This form is primarily used to
define \textit{macros}, which are functions from syntax to syntax.  The
following is a macro that always produces the constant expression
\RktVal{5}:

\noindent \begin{SCodeFlow}\RktPn{(}\RktSym{\RktStxLink{define{-}syntax}}\mbox{\hphantom{\Scribtexttt{x}}}\RktPn{(}\RktSym{always{-}5}\mbox{\hphantom{\Scribtexttt{x}}}\RktSym{stx}\RktPn{)}\mbox{\hphantom{\Scribtexttt{x}}}\RktRdr{\#{'}}\RktVal{5}\RktPn{)}\end{SCodeFlow}

\noindent \RktSym{always{-}5} is a function with formal parameter \RktSym{stx}.  When the
\RktSym{always{-}5} macro is used, this function is provided the syntax of
the use as an argument.  Macros always produce
 \textit{syntax objects}, here created with the \RktSym{\RktStxLink{syntax}}
constructor (abbreviated here \RktInBG{\RktIn{\#{'}}}).  That syntax object is, of course,
just the expression
\RktVal{5}.  Since \RktSym{always{-}5} is bound by \RktSym{\RktStxLink{define{-}syntax}}, it is a
macro, and can be used in expressions such as \RktPn{(}\RktSym{\RktValLink{+}}\Scribtexttt{ }\RktVal{3}\Scribtexttt{ }\RktPn{(}\RktSym{always{-}5}\RktPn{)}\RktPn{)}.
This expression is rewritten \textit{at compile time} to the
expression \RktPn{(}\RktSym{\RktValLink{+}}\Scribtexttt{ }\RktVal{3}\Scribtexttt{ }\RktVal{5}\RktPn{)}, which then evaluates as usual.

Of course, most syntactic extensions examine their arguments.
Typically, these are written with pattern matching to simplify
destructuring.\footnote{For more on the relationship between \RktSym{\RktStxLink{match}}
and these pattern matchers, see \ChapRefUC{5}{History and Related Work}.}  For example, the
following defines a macro which takes a function and an argument, and
adds a debugging printout of the argument value:

\noindent \begin{SCodeFlow}\begin{RktBlk}\begin{SingleColumn}\RktPn{(}\RktSym{\RktStxLink{define{-}syntax}}\mbox{\hphantom{\Scribtexttt{x}}}\RktPn{(}\RktSym{debug{-}call}\mbox{\hphantom{\Scribtexttt{x}}}\RktSym{stx}\RktPn{)}

\mbox{\hphantom{\Scribtexttt{xx}}}\RktPn{(}\RktSym{\RktStxLink{syntax{-}parse}}\mbox{\hphantom{\Scribtexttt{x}}}\RktSym{stx}

\mbox{\hphantom{\Scribtexttt{xxx}}}\RktPn{[}\RktPn{(}\RktSym{debug{-}call}\mbox{\hphantom{\Scribtexttt{x}}}\RktSym{fun}\mbox{\hphantom{\Scribtexttt{x}}}\RktSym{arg}\RktPn{)}

\mbox{\hphantom{\Scribtexttt{xxxx}}}\RktRdr{\#{'}}\RktPn{(}\RktSym{\RktStxLink{let}}\mbox{\hphantom{\Scribtexttt{x}}}\RktPn{(}\RktPn{[}\RktSym{tmp}\mbox{\hphantom{\Scribtexttt{x}}}\RktSym{arg}\RktPn{]}\RktPn{)}

\mbox{\hphantom{\Scribtexttt{xxxxxxxx}}}\RktPn{(}\RktSym{\RktValLink{printf}}\mbox{\hphantom{\Scribtexttt{x}}}\RktVal{"argument $\sim$a was{\hbox{\texttt{:}}} $\sim$a{\char`\\}n"}\mbox{\hphantom{\Scribtexttt{x}}}\RktVal{{'}}\RktVal{arg}\mbox{\hphantom{\Scribtexttt{x}}}\RktSym{tmp}\RktPn{)}

\mbox{\hphantom{\Scribtexttt{xxxxxxxx}}}\RktPn{(}\RktSym{fun}\mbox{\hphantom{\Scribtexttt{x}}}\RktSym{tmp}\RktPn{)}\RktPn{)}\RktPn{]}\RktPn{)}\RktPn{)}\end{SingleColumn}\end{RktBlk}\end{SCodeFlow}

\noindent \begin{SCodeFlow}\begin{SingleColumn}\Scribtexttt{{\texttt >} }\RktPn{(}\RktSym{debug{-}call}\mbox{\hphantom{\Scribtexttt{x}}}\RktSym{\RktValLink{add1}}\mbox{\hphantom{\Scribtexttt{x}}}\RktPn{(}\RktSym{\RktValLink{+}}\mbox{\hphantom{\Scribtexttt{x}}}\RktVal{3}\mbox{\hphantom{\Scribtexttt{x}}}\RktVal{4}\RktPn{)}\RktPn{)}

\RktOut{argument (+ 3 4) was{\hbox{\texttt{:}}} 7}

\RktRes{8}\end{SingleColumn}\end{SCodeFlow}

\noindent Here the pattern \RktPn{(}\RktSym{debug{-}call}\Scribtexttt{ }\RktSym{fun}\Scribtexttt{ }\RktSym{arg}\RktPn{)}  binds the variables \RktSym{fun}
and \RktSym{arg} which are then used in the result.

Finally, macros such as \RktSym{debug{-}call} can perform computation in
addition to pattern substitution to
determine the resulting expression.  We
use this ability to define the \RktSym{\RktStxLink{match}} syntactic extension.

\subsubsection[The Basics of \RktSym{\RktStxLink{match}}]{The Basics of \RktSym{\RktStxLink{match}}}\label{t:x28part_x22Thex5fBasicsx5fofx5fmatchx22x29}

\begin{Centerfigure}\begin{FigureInside}\begin{SCodeFlow}\begin{RktBlk}\begin{SingleColumn}\RktPn{(}\RktSym{\RktStxLink{define{-}syntax}}\mbox{\hphantom{\Scribtexttt{x}}}\RktPn{(}\RktSym{simple{-}match}\mbox{\hphantom{\Scribtexttt{x}}}\RktSym{stx}\RktPn{)}

\mbox{\hphantom{\Scribtexttt{xx}}}\RktPn{(}\RktSym{\RktStxLink{define}}\mbox{\hphantom{\Scribtexttt{x}}}\RktPn{(}\RktSym{parse{-}pat}\mbox{\hphantom{\Scribtexttt{x}}}\RktSym{pat}\mbox{\hphantom{\Scribtexttt{x}}}\RktSym{rhs}\RktPn{)}

\mbox{\hphantom{\Scribtexttt{xxxx}}}\RktPn{(}\RktSym{\RktStxLink{syntax{-}parse}}\mbox{\hphantom{\Scribtexttt{x}}}\RktSym{pat}

\mbox{\hphantom{\Scribtexttt{xxxxxx}}}\RktPn{[}\RktSym{x{\hbox{\texttt{:}}}id}

\mbox{\hphantom{\Scribtexttt{xxxxxxx}}}\RktRdr{\#{`}}\RktPn{[}\RktSym{\RktValLink{true}}\mbox{\hphantom{\Scribtexttt{xxxxxxxxxxxxxxxx}}}\RktPn{(}\RktSym{\RktStxLink{let}}\mbox{\hphantom{\Scribtexttt{x}}}\RktPn{(}\RktPn{[}\RktSym{x}\mbox{\hphantom{\Scribtexttt{x}}}\RktSym{tmp}\RktPn{]}\RktPn{)}\mbox{\hphantom{\Scribtexttt{x}}}\RktRdr{\#,}\RktSym{rhs}\RktPn{)}\RktPn{]}\RktPn{]}

\mbox{\hphantom{\Scribtexttt{xxxxxx}}}\RktPn{[}\RktVal{{'}}\RktVal{val}

\mbox{\hphantom{\Scribtexttt{xxxxxxx}}}\RktRdr{\#{`}}\RktPn{[}\RktPn{(}\RktSym{\RktValLink{equal{\hbox{\texttt{?}}}}}\mbox{\hphantom{\Scribtexttt{x}}}\RktSym{tmp}\mbox{\hphantom{\Scribtexttt{x}}}\RktVal{{'}}\RktVal{val}\RktPn{)}\mbox{\hphantom{\Scribtexttt{xxxxxxxxxxxxxxxxxx}}}\RktRdr{\#,}\RktSym{rhs}\RktPn{]}\RktPn{]}\RktPn{)}\RktPn{)}

\mbox{\hphantom{\Scribtexttt{xx}}}\RktPn{(}\RktSym{\RktStxLink{syntax{-}parse}}\mbox{\hphantom{\Scribtexttt{x}}}\RktSym{stx}

\mbox{\hphantom{\Scribtexttt{xxxx}}}\RktPn{[}\RktPn{(}\RktSym{simple{-}match}\mbox{\hphantom{\Scribtexttt{x}}}\RktSym{e{\hbox{\texttt{:}}}expr}\mbox{\hphantom{\Scribtexttt{x}}}\RktPn{[}\RktSym{pat}\mbox{\hphantom{\Scribtexttt{x}}}\RktSym{rhs}\RktPn{]}\mbox{\hphantom{\Scribtexttt{x}}}\RktSym{\RktStxLink{{\hbox{\texttt{.}}}{\hbox{\texttt{.}}}{\hbox{\texttt{.}}}}}\RktPn{)}

\mbox{\hphantom{\Scribtexttt{xxxxx}}}\RktPn{(}\RktSym{\RktStxLink{with{-}syntax}}\mbox{\hphantom{\Scribtexttt{x}}}\RktPn{(}\RktPn{[}\RktPn{(}\RktSym{new{-}clause}\mbox{\hphantom{\Scribtexttt{x}}}\RktSym{\RktStxLink{{\hbox{\texttt{.}}}{\hbox{\texttt{.}}}{\hbox{\texttt{.}}}}}\RktPn{)}

\mbox{\hphantom{\Scribtexttt{xxxxxxxxxxxxxxxxxxxx}}}\RktPn{(}\RktSym{\RktValLink{syntax{-}map}}\mbox{\hphantom{\Scribtexttt{x}}}\RktSym{parse{-}pat}

\mbox{\hphantom{\Scribtexttt{xxxxxxxxxxxxxxxxxxxxxxxxxxxxxxxx}}}\RktRdr{\#{'}}\RktPn{(}\RktSym{pat}\mbox{\hphantom{\Scribtexttt{x}}}\RktSym{\RktStxLink{{\hbox{\texttt{.}}}{\hbox{\texttt{.}}}{\hbox{\texttt{.}}}}}\RktPn{)}\mbox{\hphantom{\Scribtexttt{x}}}\RktRdr{\#{'}}\RktPn{(}\RktSym{rhs}\mbox{\hphantom{\Scribtexttt{x}}}\RktSym{\RktStxLink{{\hbox{\texttt{.}}}{\hbox{\texttt{.}}}{\hbox{\texttt{.}}}}}\RktPn{)}\RktPn{)}\RktPn{]}\RktPn{)}

\mbox{\hphantom{\Scribtexttt{xxxxxxx}}}\RktRdr{\#{'}}\RktPn{(}\RktSym{\RktStxLink{let}}\mbox{\hphantom{\Scribtexttt{x}}}\RktPn{(}\RktPn{[}\RktSym{tmp}\mbox{\hphantom{\Scribtexttt{x}}}\RktSym{e}\RktPn{]}\RktPn{)}

\mbox{\hphantom{\Scribtexttt{xxxxxxxxxxx}}}\RktPn{(}\RktSym{\RktStxLink{cond}}\mbox{\hphantom{\Scribtexttt{x}}}\RktSym{new{-}clause}\mbox{\hphantom{\Scribtexttt{x}}}\RktSym{\RktStxLink{{\hbox{\texttt{.}}}{\hbox{\texttt{.}}}{\hbox{\texttt{.}}}}}

\mbox{\hphantom{\Scribtexttt{xxxxxxxxxxxxxxxxx}}}\RktPn{[}\RktSym{\RktStxLink{else}}\mbox{\hphantom{\Scribtexttt{x}}}\RktPn{(}\RktSym{\RktValLink{error}}\mbox{\hphantom{\Scribtexttt{x}}}\RktVal{"match failed"}\RktPn{)}\RktPn{]}\RktPn{)}\RktPn{)}\RktPn{)}\RktPn{]}\RktPn{)}\RktPn{)}\end{SingleColumn}\end{RktBlk}\end{SCodeFlow}

\Legend{\label{t:x28counter_x28x22figurex22_x22simplex2dmatchx22x29x29}Figure~2: A simple pattern matcher}\end{FigureInside}\end{Centerfigure}

Implementing complex language extensions such as \RktSym{\RktStxLink{match}} requires more care
than extensions that are expressed as simple rewrite rules.
A simple pattern matcher supporting only variable and literal patterns
is given in Figure~2.  The basic architecture is as
follows:

\noindent \begin{itemize}\atItemizeStart

\item The expression being matched, \RktSym{e}, is bound to a new temporary
variable, \RktSym{tmp}.

\item Each clause, consisting of a pattern \RktSym{pat} and a right{-}hand side \RktSym{rhs},
 is transformed into a clause suitable for \RktSym{\RktStxLink{cond}}
by the \RktSym{parse{-}pat} function.
For literal patterns, the test expression is an equality check against
\RktSym{tmp}.  For variable patterns, the test is trivial, but the
variable from the pattern is bound to \RktSym{tmp} in the right{-}hand side.

\item Finally, all of the clauses are placed in a single \RktSym{\RktStxLink{cond}}
expression with an \RktSym{\RktStxLink{else}} clause that throws a pattern{-}match failure.\end{itemize}

\noindent This matcher is missing most of the features of \RktSym{\RktStxLink{match}}, but demonstrates
the key aspects of defining \RktSym{\RktStxLink{match}} as a syntactic extension.  In the
subsequent sections, we will see how to extend this matcher via the
same syntactic extension framework in which it is implemented.

\subsubsection[A Production Implementation]{A Production Implementation}\label{t:x28part_x22Ax5fProductionx5fImplementationx22x29}

  The \RktSym{\RktStxLink{match}}
implementation first translates each pattern into an abstract syntax
tree of patterns while also simplifying patterns to remove
redundancy.  For example, \RktKw{list} patterns are simplified to
\RktKw{cons} patterns, and patterns with implicit binding are
rewritten to use \RktSym{\RktStxLink{and}} with variable patterns.

Given a table of pattern structures, both column
optimizations~\cite{maranget} and row
optimizations~\cite{maranget-tree} are performed to select the best
order to for matching patterns and to coalesce related patterns.  Finally,
the Augustsson algorithm is used to generate the residual code.

\sectionNewpage

\section[Advanced Patterns]{Advanced Patterns}\label{t:x28part_x22advancedx22x29}

Extending the simple patterns described in \ChapRefUC{2}{Pattern Matching in Racket}, we now
add repetition to lists, patterns that transform their input,
 and matching of user{-}defined structures.
 These patterns are necessary  both to support the
examples presented in \ChapRefUC{1}{Extending Pattern Matching},  as well as introducing
concepts that are used in the definition of \RktSym{\RktStxLink{match}} expanders.

\subsection[Lists with Repetition]{Lists with Repetition}\label{t:x28part_x22Listsx5fwithx5fRepetitionx22x29}

The first significant extension is the ability to describe
arbitrary{-}length lists.  Of course, we can already describe
some uses of such lists using the \RktSym{\RktValLink{cons}} pattern:

\begin{SCodeFlow}\begin{RktBlk}\begin{SingleColumn}\RktPn{(}\RktSym{\RktStxLink{match}}\mbox{\hphantom{\Scribtexttt{x}}}\RktPn{(}\RktSym{\RktValLink{list}}\mbox{\hphantom{\Scribtexttt{x}}}\RktVal{1}\mbox{\hphantom{\Scribtexttt{x}}}\RktVal{2}\mbox{\hphantom{\Scribtexttt{x}}}\RktVal{3}\RktPn{)}

\mbox{\hphantom{\Scribtexttt{x}}}\RktPn{[}\RktPn{(}\RktSym{\RktValLink{cons}}\mbox{\hphantom{\Scribtexttt{x}}}\RktPn{(}\RktSym{\RktStxLink{and}}\mbox{\hphantom{\Scribtexttt{x}}}\RktSym{x}\mbox{\hphantom{\Scribtexttt{x}}}\RktPn{(}\RktKw{{\hbox{\texttt{?}}}}\mbox{\hphantom{\Scribtexttt{x}}}\RktSym{\RktValLink{number{\hbox{\texttt{?}}}}}\RktPn{)}\RktPn{)}\mbox{\hphantom{\Scribtexttt{x}}}\RktSym{xs}\RktPn{)}\mbox{\hphantom{\Scribtexttt{x}}}\RktSym{x}\RktPn{]}\RktPn{)}\end{SingleColumn}\end{RktBlk}\end{SCodeFlow}

However, the use of \RktSym{\RktStxLink{{\hbox{\texttt{.}}}{\hbox{\texttt{.}}}{\hbox{\texttt{.}}}}} allows the specification of
arbitrary{-}length lists with a specification of a pattern to be matched
against each element. For example, to match a list of numbers:

\begin{SCodeFlow}\begin{RktBlk}\begin{SingleColumn}\RktPn{(}\RktSym{\RktStxLink{match}}\mbox{\hphantom{\Scribtexttt{x}}}\RktPn{(}\RktSym{\RktValLink{list}}\mbox{\hphantom{\Scribtexttt{x}}}\RktVal{1}\mbox{\hphantom{\Scribtexttt{x}}}\RktVal{2}\mbox{\hphantom{\Scribtexttt{x}}}\RktVal{3}\RktPn{)}

\mbox{\hphantom{\Scribtexttt{x}}}\RktPn{[}\RktPn{(}\RktSym{\RktValLink{list}}\mbox{\hphantom{\Scribtexttt{x}}}\RktPn{(}\RktSym{\RktStxLink{and}}\mbox{\hphantom{\Scribtexttt{x}}}\RktSym{xs}\mbox{\hphantom{\Scribtexttt{x}}}\RktPn{(}\RktKw{{\hbox{\texttt{?}}}}\mbox{\hphantom{\Scribtexttt{x}}}\RktSym{\RktValLink{number{\hbox{\texttt{?}}}}}\RktPn{)}\RktPn{)}\mbox{\hphantom{\Scribtexttt{x}}}\RktSym{\RktStxLink{{\hbox{\texttt{.}}}{\hbox{\texttt{.}}}{\hbox{\texttt{.}}}}}\RktPn{)}\mbox{\hphantom{\Scribtexttt{x}}}\RktSym{xs}\RktPn{]}\RktPn{)}\end{SingleColumn}\end{RktBlk}\end{SCodeFlow}

\noindent This checks that every element of the list is a number, and binds
\RktSym{xs} to a \textit{list} of numbers (or fails to match).
The \RktSym{\RktStxLink{{\hbox{\texttt{.}}}{\hbox{\texttt{.}}}{\hbox{\texttt{.}}}}} suffix functions similarly to the Kleene closure operator in
regular expressions.  In fact, in conjunction with \RktSym{\RktStxLink{or}} and \RktSym{\RktStxLink{and}}
patterns \RktSym{\RktStxLink{match}} patterns can express many regular languages.

\subsubsection[Compilation]{Compilation}\label{t:x28part_x22Compilationx22x29}

Compiling patterns with \RktSym{\RktStxLink{{\hbox{\texttt{.}}}{\hbox{\texttt{.}}}{\hbox{\texttt{.}}}}} requires a straightforward extension
to the traditional pattern matching compilation algorithm.  A pattern clause
such as \RktPn{[}\RktPn{(}\RktSym{\RktValLink{list}}\Scribtexttt{ }\RktSym{p}\Scribtexttt{ }\RktSym{\RktStxLink{{\hbox{\texttt{.}}}{\hbox{\texttt{.}}}{\hbox{\texttt{.}}}}}\RktPn{)}\Scribtexttt{ }\RktSym{\RktStxLink{expr}}\RktPn{]} is compiled using two clauses.  The
first is \RktPn{[}\RktPn{(}\RktSym{\RktValLink{list}}\RktPn{)}\Scribtexttt{ }\RktSym{\RktStxLink{expr}}\RktPn{]}, matching empty list and the second is
\RktPn{[}\RktPn{(}\RktSym{\RktValLink{cons}}\Scribtexttt{ }\RktSym{p}\Scribtexttt{ }\RktSym{\RktValLink{rest}}\RktPn{)}\Scribtexttt{ }\RktPn{(}\RktSym{loop}\Scribtexttt{ }\RktSym{\RktValLink{rest}}\RktPn{)}\RktPn{]} where \RktSym{loop} is a recursive function
that repeats the matching again on its argument.  Of course, all of
the  variables bound by \RktSym{p} must be passed along in the loop so that
they are accumulated as a list.

\subsection[\RktSym{\badlink{\RktValLink{app}}} Patterns]{\RktSym{\badlink{\RktValLink{app}}} Patterns}\label{t:x28part_x22appx5fPatternsx22x29}

The simplest form of extensible pattern matching is the \RktSym{\badlink{\RktValLink{app}}}
pattern.  A pattern of the form \RktPn{(}\RktSym{\badlink{\RktValLink{app}}}\Scribtexttt{ }\RktSym{f}\Scribtexttt{ }\RktSym{pat}\RktPn{)} applies the
(user{-}specified) function  \RktSym{f} to the value being matched, and then
matches \RktSym{pat} against the \textit{result} of \RktSym{f}. For example:

\noindent \begin{SCodeFlow}\begin{SingleColumn}\begin{RktBlk}\begin{SingleColumn}\Scribtexttt{{\texttt >} }\RktPn{(}\RktSym{\RktStxLink{match}}\mbox{\hphantom{\Scribtexttt{x}}}\RktVal{16}

\RktPlain{\mbox{\hphantom{\Scribtexttt{xx}}}}\mbox{\hphantom{\Scribtexttt{x}}}\RktPn{[}\RktPn{(}\RktSym{\badlink{\RktValLink{app}}}\mbox{\hphantom{\Scribtexttt{x}}}\RktSym{\RktValLink{sqrt}}\mbox{\hphantom{\Scribtexttt{x}}}\RktPn{(}\RktSym{\RktStxLink{and}}\mbox{\hphantom{\Scribtexttt{x}}}\RktPn{(}\RktKw{{\hbox{\texttt{?}}}}\mbox{\hphantom{\Scribtexttt{x}}}\RktSym{\RktValLink{integer{\hbox{\texttt{?}}}}}\RktPn{)}\mbox{\hphantom{\Scribtexttt{x}}}\RktSym{x}\RktPn{)}\RktPn{)}

\RktPlain{\mbox{\hphantom{\Scribtexttt{xx}}}}\mbox{\hphantom{\Scribtexttt{xx}}}\RktPn{(}\RktSym{\RktValLink{format}}\mbox{\hphantom{\Scribtexttt{x}}}\RktVal{"perfect square{\hbox{\texttt{:}}} $\sim$a"}\mbox{\hphantom{\Scribtexttt{x}}}\RktSym{x}\RktPn{)}\RktPn{]}

\RktPlain{\mbox{\hphantom{\Scribtexttt{xx}}}}\mbox{\hphantom{\Scribtexttt{x}}}\RktPn{[}\RktSym{\RktStxLink{{\char`\_}}}\mbox{\hphantom{\Scribtexttt{x}}}\RktVal{"not perfect"}\RktPn{]}\RktPn{)}\end{SingleColumn}\end{RktBlk}

\RktRes{"perfect square{\hbox{\texttt{:}}} 4"}\end{SingleColumn}\end{SCodeFlow}

\noindent Implementing the \RktSym{\badlink{\RktValLink{app}}} pattern is straightforward.  We simply apply
the function to the value being matched, and continue pattern matching
with the new result and the sub{-}pattern.  This fits straightforwardly
into the Augustsson{-}style matching algorithm.  Currently, we do not
attempt to coalesce multiple uses of the same expression, however this
would be a straightforward addition.  As a result of the backtracking
algorithm, the function may be called multiple times; the use of
 side{-}effects in this function is therefore discouraged.

\RktSym{\badlink{\RktValLink{app}}} patterns are already very expressive{---}Peyton
Jones~\cite{spj:view-patterns} gives many examples of their use under
the name \textit{view patterns}. Below is one simple demonstration of their
use:

\noindent \begin{SCodeFlow}\begin{SingleColumn}\begin{RktBlk}\begin{SingleColumn}\RktPn{(}\RktSym{\RktStxLink{define}}\mbox{\hphantom{\Scribtexttt{x}}}\RktPn{(}\RktSym{\RktValLink{map}}\mbox{\hphantom{\Scribtexttt{x}}}\RktSym{f}\mbox{\hphantom{\Scribtexttt{x}}}\RktSym{l}\RktPn{)}

\mbox{\hphantom{\Scribtexttt{xx}}}\RktPn{(}\RktSym{\RktStxLink{match}}\mbox{\hphantom{\Scribtexttt{x}}}\RktSym{l}

\mbox{\hphantom{\Scribtexttt{xxxx}}}\RktPn{[}\RktVal{{'}}\RktVal{(}\RktVal{)}\mbox{\hphantom{\Scribtexttt{x}}}\RktVal{{'}}\RktVal{(}\RktVal{)}\RktPn{]}

\mbox{\hphantom{\Scribtexttt{xxxx}}}\RktPn{[}\RktPn{(}\RktSym{\RktValLink{cons}}\mbox{\hphantom{\Scribtexttt{x}}}\RktPn{(}\RktSym{\badlink{\RktValLink{app}}}\mbox{\hphantom{\Scribtexttt{x}}}\RktSym{f}\mbox{\hphantom{\Scribtexttt{x}}}\RktSym{x}\RktPn{)}\mbox{\hphantom{\Scribtexttt{x}}}\RktPn{(}\RktSym{\badlink{\RktValLink{app}}}\mbox{\hphantom{\Scribtexttt{x}}}\RktPn{(}\RktSym{\RktValLink{curry}}\mbox{\hphantom{\Scribtexttt{x}}}\RktSym{\RktValLink{map}}\mbox{\hphantom{\Scribtexttt{x}}}\RktSym{f}\RktPn{)}\mbox{\hphantom{\Scribtexttt{x}}}\RktSym{xs}\RktPn{)}\RktPn{)}

\mbox{\hphantom{\Scribtexttt{xxxxx}}}\RktPn{(}\RktSym{\RktValLink{cons}}\mbox{\hphantom{\Scribtexttt{x}}}\RktSym{x}\mbox{\hphantom{\Scribtexttt{x}}}\RktSym{xs}\RktPn{)}\RktPn{]}\RktPn{)}\RktPn{)}\end{SingleColumn}\end{RktBlk}

~

\begin{SingleColumn}\Scribtexttt{{\texttt >} }\RktPn{(}\RktSym{\RktValLink{map}}\mbox{\hphantom{\Scribtexttt{x}}}\RktSym{\RktValLink{add1}}\mbox{\hphantom{\Scribtexttt{x}}}\RktPn{(}\RktSym{\RktValLink{list}}\mbox{\hphantom{\Scribtexttt{x}}}\RktVal{1}\mbox{\hphantom{\Scribtexttt{x}}}\RktVal{2}\mbox{\hphantom{\Scribtexttt{x}}}\RktVal{3}\mbox{\hphantom{\Scribtexttt{x}}}\RktVal{4}\mbox{\hphantom{\Scribtexttt{x}}}\RktVal{5}\RktPn{)}\RktPn{)}

\RktRes{{'}(2\Scribtexttt{ }3\Scribtexttt{ }4\Scribtexttt{ }5\Scribtexttt{ }6)}\end{SingleColumn}\end{SingleColumn}\end{SCodeFlow}

\noindent In combination with the ability to define new pattern forms, \RktSym{\badlink{\RktValLink{app}}}
patterns allow almost arbitrary extensions to patterns.

\subsection[User{-}defined Structures]{User{-}defined Structures}\label{t:x28part_x22structx22x29}

Racket supports the definition of new user{-}specified structures:

\begin{SCodeFlow}\begin{RktBlk}\begin{SingleColumn}\RktPn{(}\RktKw{struct}\mbox{\hphantom{\Scribtexttt{x}}}\RktSym{point}\mbox{\hphantom{\Scribtexttt{x}}}\RktPn{(}\RktSym{x}\mbox{\hphantom{\Scribtexttt{x}}}\RktSym{y}\RktPn{)}\RktPn{)}

\RktPn{(}\RktSym{\RktStxLink{define}}\mbox{\hphantom{\Scribtexttt{x}}}\RktSym{p1}\mbox{\hphantom{\Scribtexttt{x}}}\RktPn{(}\RktSym{make{-}point}\mbox{\hphantom{\Scribtexttt{x}}}\RktVal{1}\mbox{\hphantom{\Scribtexttt{x}}}\RktVal{2}\RktPn{)}\RktPn{)}\end{SingleColumn}\end{RktBlk}\end{SCodeFlow}

\noindent \begin{SCodeFlow}\begin{SingleColumn}\Scribtexttt{{\texttt >} }\RktPn{(}\RktSym{point{-}x}\mbox{\hphantom{\Scribtexttt{x}}}\RktSym{p1}\RktPn{)}

\RktRes{1}\end{SingleColumn}\end{SCodeFlow}

\noindent Of course, they should also be supported in pattern matching:

\noindent \begin{SCodeFlow}\begin{SingleColumn}\begin{RktBlk}\begin{SingleColumn}\Scribtexttt{{\texttt >} }\RktPn{(}\RktSym{\RktStxLink{match}}\mbox{\hphantom{\Scribtexttt{x}}}\RktSym{p1}

\RktPlain{\mbox{\hphantom{\Scribtexttt{xx}}}}\mbox{\hphantom{\Scribtexttt{x}}}\RktPn{[}\RktPn{(}\RktSym{point}\mbox{\hphantom{\Scribtexttt{x}}}\RktSym{a}\mbox{\hphantom{\Scribtexttt{x}}}\RktSym{b}\RktPn{)}\mbox{\hphantom{\Scribtexttt{x}}}\RktSym{a}\RktPn{]}\RktPn{)}\end{SingleColumn}\end{RktBlk}

\RktRes{1}\end{SingleColumn}\end{SCodeFlow}

\noindent To accomplish this, the \RktKw{struct} form, which is also implemented as
a language extension, must communicate with
\RktSym{\RktStxLink{match}} \textit{at expansion time}, so that \RktSym{\RktStxLink{match}} can produce code using the
\RktSym{point{\hbox{\texttt{?}}}} predicate and \RktSym{point{-}x} selector.

\subsubsection[Static Binding]{Static Binding}\label{t:x28part_x22Staticx5fBindingx22x29}

The simplest form of communication between two portions of the program
is binding.  To take advantage of this at expansion time, we can
 bind arbitrary values with the \RktSym{\RktStxLink{define{-}syntax}} form, not just macros.

\noindent \begin{SCodeFlow}\RktPn{(}\RktSym{\RktStxLink{define{-}syntax}}\mbox{\hphantom{\Scribtexttt{x}}}\RktSym{just{-}5}\mbox{\hphantom{\Scribtexttt{x}}}\RktVal{5}\RktPn{)}\end{SCodeFlow}

\noindent Now, \RktSym{just{-}5} is bound to \RktVal{5} in the static environment.  We can
access this environment with the \RktSym{\RktValLink{syntax{-}local{-}value}} function,
rewriting the \RktSym{always{-}5} macro thus:

\noindent \begin{SCodeFlow}\begin{RktBlk}\begin{SingleColumn}\RktPn{(}\RktSym{\RktStxLink{define{-}syntax}}\mbox{\hphantom{\Scribtexttt{x}}}\RktPn{(}\RktSym{always{-}5}\mbox{\hphantom{\Scribtexttt{x}}}\RktSym{stx}\RktPn{)}

\mbox{\hphantom{\Scribtexttt{xx}}}\RktPn{(}\RktSym{to{-}syntax}\mbox{\hphantom{\Scribtexttt{x}}}\RktPn{(}\RktSym{\RktValLink{syntax{-}local{-}value}}\mbox{\hphantom{\Scribtexttt{x}}}\RktRdr{\#{'}}\RktSym{just{-}5}\RktPn{)}\RktPn{)}\RktPn{)}\end{SingleColumn}\end{RktBlk}\end{SCodeFlow}

\begin{SCodeFlow}\begin{SingleColumn}\Scribtexttt{{\texttt >} }\RktPn{(}\RktSym{always{-}5}\RktPn{)}

\RktErr{reference to undefined identifier: to{-}syntax}\end{SingleColumn}\end{SCodeFlow}

\noindent \RktSym{\RktValLink{syntax{-}local{-}value}} produces the value to which an identifier such
as \RktRdr{\#{'}}\RktSym{just{-}5} is bound.  We then convert that value to a piece of
syntax with \RktSym{to{-}syntax}.  

\subsubsection[Static Structure Information]{Static Structure Information}\label{t:x28part_x22Staticx5fStructurex5fInformationx22x29}

We can take advantage of this static binding mechanism to record
statically known information about structure definitions.  The
\RktPn{(}\RktKw{struct}\Scribtexttt{ }\RktSym{point}\Scribtexttt{ }\RktPn{(}\RktSym{x}\Scribtexttt{ }\RktSym{y}\RktPn{)}\RktPn{)} form expands to definitions of the runtime values:

\noindent \begin{SCodeFlow}\RktPn{(}\RktSym{\RktStxLink{define}}\mbox{\hphantom{\Scribtexttt{x}}}\RktSym{point{-}x}\mbox{\hphantom{\Scribtexttt{x}}}\relax{---}\RktPn{)}\mbox{\hphantom{\Scribtexttt{x}}}\RktPn{(}\RktSym{\RktStxLink{define}}\mbox{\hphantom{\Scribtexttt{x}}}\RktSym{point{-}y}\mbox{\hphantom{\Scribtexttt{x}}}\relax{---}\RktPn{)}\mbox{\hphantom{\Scribtexttt{x}}}\RktPn{(}\RktSym{\RktStxLink{define}}\mbox{\hphantom{\Scribtexttt{x}}}\RktSym{make{-}point}\mbox{\hphantom{\Scribtexttt{x}}}\relax{---}\RktPn{)}\end{SCodeFlow}

\noindent as well as \textit{compile{-}time} static information about the structure.

\noindent \begin{SCodeFlow}\begin{RktBlk}\begin{SingleColumn}\RktPn{(}\RktSym{\RktStxLink{define{-}syntax}}\mbox{\hphantom{\Scribtexttt{x}}}\RktSym{point}

\mbox{\hphantom{\Scribtexttt{xx}}}\RktPn{(}\RktSym{\RktValLink{list}}\mbox{\hphantom{\Scribtexttt{x}}}\RktRdr{\#{'}}\RktSym{make{-}point}\mbox{\hphantom{\Scribtexttt{x}}}\RktRdr{\#{'}}\RktSym{point{\hbox{\texttt{?}}}}\mbox{\hphantom{\Scribtexttt{x}}}\RktRdr{\#{'}}\RktSym{point{-}x}\mbox{\hphantom{\Scribtexttt{x}}}\RktRdr{\#{'}}\RktSym{point{-}y}\RktPn{)}\RktPn{)}\end{SingleColumn}\end{RktBlk}\end{SCodeFlow}

\noindent Thus, the identifier \RktSym{point} contains information about how to
create \RktSym{point}s, test for them, and extract their components.
Using this facility, we can now extend our simple pattern matcher from
Figure~2  to test for structures.  We add a new
clause to the \RktSym{parse{-}pat} function:

\noindent \begin{SCodeFlow}\begin{RktBlk}\begin{SingleColumn}\RktPn{[}\RktPn{(}\RktKw{struct}\mbox{\hphantom{\Scribtexttt{x}}}\RktSym{sname}\RktPn{)}

\mbox{\hphantom{\Scribtexttt{xx}}}\RktPn{(}\RktSym{\RktStxLink{let*}}\mbox{\hphantom{\Scribtexttt{x}}}\RktPn{(}\RktPn{[}\RktSym{static{-}info}\mbox{\hphantom{\Scribtexttt{x}}}\RktPn{(}\RktSym{\RktValLink{syntax{-}local{-}value}}\mbox{\hphantom{\Scribtexttt{x}}}\RktRdr{\#{'}}\RktSym{sname}\RktPn{)}\RktPn{]}

\mbox{\hphantom{\Scribtexttt{xxxxxxxxx}}}\RktPn{[}\RktSym{predicate}\mbox{\hphantom{\Scribtexttt{x}}}\RktPn{(}\RktSym{\RktValLink{second}}\mbox{\hphantom{\Scribtexttt{x}}}\RktSym{static{-}info}\RktPn{)}\RktPn{]}\RktPn{)}

\mbox{\hphantom{\Scribtexttt{xxx}}}\RktRdr{\#{`}}\RktPn{[}\RktPn{(}\RktRdr{\#,}\RktSym{predicate}\mbox{\hphantom{\Scribtexttt{x}}}\RktSym{tmp}\RktPn{)}\mbox{\hphantom{\Scribtexttt{xxxxxxxxxxxxxxxxxxxx}}}\RktRdr{\#,}\RktSym{rhs}\RktPn{]}\RktPn{)}\RktPn{]}\end{SingleColumn}\end{RktBlk}\end{SCodeFlow}

\noindent In this, we first access the static structure information about the
mentioned structure (\RktSym{static{-}info}), and select the element naming
the structure predicate (\RktSym{predicate}).  Then the result clause
simply uses the predicate to test \RktSym{tmp}.

In Racket, of course, the \RktKw{struct} pattern also takes patterns to
match against each field of the structure.  Compiling these patterns
uses the field accessors names provided by the static information.

\sectionNewpage

\section[Extensible Patterns]{Extensible Patterns}\label{t:x28part_x22extensiblex22x29}

Using the techniques of  \SecRefUC{3.3}{User{-}defined Structures}, we now add extensibility
to \RktSym{\RktStxLink{match}}.

\subsection[\RktSym{\RktStxLink{match}} Expanders]{\RktSym{\RktStxLink{match}} Expanders}\label{t:x28part_x22matchx5fExpandersx22x29}

The first task is to define the API for extending \RktSym{\RktStxLink{match}}.  The
fundamental mechanism is a procedure that consumes
the syntax for a pattern using the extension and produces syntax for a
new pattern.  For example, the
following procedure always produces a pattern matching numbers:

\noindent \begin{SCodeFlow}\RktPn{(}\RktSym{\RktStxLink{$\lambda$}}\mbox{\hphantom{\Scribtexttt{x}}}\RktPn{(}\RktSym{stx}\RktPn{)}\mbox{\hphantom{\Scribtexttt{x}}}\RktRdr{\#{'}}\RktPn{(}\RktKw{{\hbox{\texttt{?}}}}\mbox{\hphantom{\Scribtexttt{x}}}\RktSym{\RktValLink{number{\hbox{\texttt{?}}}}}\RktPn{)}\RktPn{)}\end{SCodeFlow}

Of course, more interesting functions are possible.  This example combines the
\RktSym{\RktStxLink{and}} and \RktKw{{\hbox{\texttt{?}}}} pattern to make them more useful together than the
ones presented in our original grammar in
Figure~1:\footnote{Racket{'}s version of \RktSym{\RktStxLink{match}}
includes this functionality in the \RktKw{{\hbox{\texttt{?}}}} pattern, as demonstrated in the first
section.}

\noindent \begin{SCodeFlow}\begin{RktBlk}\begin{SingleColumn}\RktPn{(}\RktSym{\RktStxLink{$\lambda$}}\mbox{\hphantom{\Scribtexttt{x}}}\RktPn{(}\RktSym{stx}\RktPn{)}

\mbox{\hphantom{\Scribtexttt{xx}}}\RktPn{(}\RktSym{\RktStxLink{syntax{-}parse}}\mbox{\hphantom{\Scribtexttt{x}}}\RktSym{stx}

\mbox{\hphantom{\Scribtexttt{xxxx}}}\RktPn{[}\RktPn{(}\RktSym{{\hbox{\texttt{?}}}{\hbox{\texttt{?}}}}\mbox{\hphantom{\Scribtexttt{x}}}\RktSym{pred}\mbox{\hphantom{\Scribtexttt{x}}}\RktSym{pat}\RktPn{)}\mbox{\hphantom{\Scribtexttt{x}}}\RktRdr{\#{'}}\RktPn{(}\RktSym{\RktStxLink{and}}\mbox{\hphantom{\Scribtexttt{x}}}\RktPn{(}\RktKw{{\hbox{\texttt{?}}}}\mbox{\hphantom{\Scribtexttt{x}}}\RktSym{pred}\RktPn{)}\mbox{\hphantom{\Scribtexttt{x}}}\RktSym{pat}\RktPn{)}\RktPn{]}\RktPn{)}\RktPn{)}\end{SingleColumn}\end{RktBlk}\end{SCodeFlow}

To inform \RktSym{\RktStxLink{match}} that this is to be used as an
extension, we introduce the \RktSym{\RktStxLink{define{-}match{-}expander}} form.
This form expects a procedure for transforming patterns, and binds it
statically.  We could simply use \RktSym{\RktStxLink{define{-}syntax}}, but this has
several drawbacks: (a) it prevents us from distinguishing \RktSym{\RktStxLink{match}} expanders
from other forms and (b) it prevents us from adding additional
information to \RktSym{\RktStxLink{match}} expanders.  For example, the Racket \RktSym{\RktStxLink{match}}
implementation supports an old{-}style syntax for
backwards{-}compatibility, and the distinction between \RktSym{\RktStxLink{match}} expanders and
other static binding enables the definition of expanders that work
with both syntaxes.

We can therefore define our trivial \RktSym{\RktStxLink{match}} expander for numbers with

\noindent \begin{SCodeFlow}\begin{RktBlk}\begin{SingleColumn}\RktPn{(}\RktSym{\RktStxLink{define{-}match{-}expander}}\mbox{\hphantom{\Scribtexttt{x}}}\RktSym{num}

\mbox{\hphantom{\Scribtexttt{xx}}}\RktPn{(}\RktSym{\RktStxLink{$\lambda$}}\mbox{\hphantom{\Scribtexttt{x}}}\RktPn{(}\RktSym{stx}\RktPn{)}\mbox{\hphantom{\Scribtexttt{x}}}\RktRdr{\#{'}}\RktPn{(}\RktKw{{\hbox{\texttt{?}}}}\mbox{\hphantom{\Scribtexttt{x}}}\RktSym{\RktValLink{number{\hbox{\texttt{?}}}}}\RktPn{)}\RktPn{)}\RktPn{)}\end{SingleColumn}\end{RktBlk}\end{SCodeFlow}

\noindent which is equivalent to

\noindent \begin{SCodeFlow}\begin{RktBlk}\begin{SingleColumn}\RktPn{(}\RktSym{\RktStxLink{define{-}syntax}}\mbox{\hphantom{\Scribtexttt{x}}}\RktSym{num}

\mbox{\hphantom{\Scribtexttt{xx}}}\RktPn{(}\RktSym{make{-}match{-}expander}\mbox{\hphantom{\Scribtexttt{x}}}\RktPn{(}\RktSym{\RktStxLink{$\lambda$}}\mbox{\hphantom{\Scribtexttt{x}}}\RktPn{(}\RktSym{stx}\RktPn{)}\mbox{\hphantom{\Scribtexttt{x}}}\RktRdr{\#{'}}\RktPn{(}\RktKw{{\hbox{\texttt{?}}}}\mbox{\hphantom{\Scribtexttt{x}}}\RktSym{\RktValLink{number{\hbox{\texttt{?}}}}}\RktPn{)}\RktPn{)}\RktPn{)}\RktPn{)}\end{SingleColumn}\end{RktBlk}\end{SCodeFlow}

\noindent This simply wraps the given procedure in a structure that \RktSym{\RktStxLink{match}} will
later recognize.
We can now  use the \RktSym{\RktStxLink{match}} expander in a pattern as follows:

\noindent \begin{SCodeFlow}\begin{SingleColumn}\begin{RktBlk}\begin{SingleColumn}\Scribtexttt{{\texttt >} }\RktPn{(}\RktSym{\RktStxLink{match}}\mbox{\hphantom{\Scribtexttt{x}}}\RktVal{7}

\RktPlain{\mbox{\hphantom{\Scribtexttt{xx}}}}\mbox{\hphantom{\Scribtexttt{x}}}\RktPn{[}\RktPn{(}\RktSym{num}\RktPn{)}\mbox{\hphantom{\Scribtexttt{x}}}\RktVal{{'}}\RktVal{yes}\RktPn{]}

\RktPlain{\mbox{\hphantom{\Scribtexttt{xx}}}}\mbox{\hphantom{\Scribtexttt{x}}}\RktPn{[}\RktSym{\RktStxLink{{\char`\_}}}\mbox{\hphantom{\Scribtexttt{x}}}\RktVal{{'}}\RktVal{no}\RktPn{]}\RktPn{)}\end{SingleColumn}\end{RktBlk}

\RktRes{{'}yes}\end{SingleColumn}\end{SCodeFlow}

\subsection[Parsing \RktSym{\RktStxLink{match}} expanders]{Parsing \RktSym{\RktStxLink{match}} expanders}\label{t:x28part_x22Parsingx5fmatchx5fexpandersx22x29}

Extending the \RktSym{parse{-}pat} function of Figure~2 to
handle \RktSym{\RktStxLink{match}} expanders is surprisingly straightforward.  We add one new
clause:

\noindent \begin{SCodeFlow}\begin{RktBlk}\begin{SingleColumn}\RktPn{[}\RktPn{(}\RktSym{expander}\mbox{\hphantom{\Scribtexttt{x}}}\RktPn{{\hbox{\texttt{.}}} }\RktSym{rest}\RktPn{)}

\mbox{\hphantom{\Scribtexttt{x}}}\RktCmt{;}\RktCmt{~}\RktCmt{(1) look up the match expander}

\mbox{\hphantom{\Scribtexttt{x}}}\RktPn{(}\RktSym{\RktStxLink{let}}\mbox{\hphantom{\Scribtexttt{x}}}\RktPn{(}\RktPn{[}\RktSym{val}\mbox{\hphantom{\Scribtexttt{x}}}\RktPn{(}\RktSym{\RktValLink{syntax{-}local{-}value}}\mbox{\hphantom{\Scribtexttt{x}}}\RktRdr{\#{'}}\RktSym{expander}\RktPn{)}\RktPn{]}\RktPn{)}

\mbox{\hphantom{\Scribtexttt{xxx}}}\RktPn{(}\RktSym{\RktStxLink{if}}\mbox{\hphantom{\Scribtexttt{x}}}\RktPn{(}\RktSym{match{-}expander{\hbox{\texttt{?}}}}\mbox{\hphantom{\Scribtexttt{x}}}\RktSym{val}\RktPn{)}

\mbox{\hphantom{\Scribtexttt{xxxxxxx}}}\RktPn{(}\RktSym{\RktStxLink{let}}\mbox{\hphantom{\Scribtexttt{x}}}\RktPn{(}\RktCmt{;}\RktCmt{~}\RktCmt{(2) extract the transformer function}

\mbox{\hphantom{\Scribtexttt{xxxxxxxxxxxxx}}}\RktPn{[}\RktSym{transformer}\mbox{\hphantom{\Scribtexttt{x}}}\RktPn{(}\RktSym{match{-}expander{-}transformer}\mbox{\hphantom{\Scribtexttt{x}}}\RktSym{val}\RktPn{)}\RktPn{]}

\mbox{\hphantom{\Scribtexttt{xxxxxxxxxxxxx}}}\RktCmt{;}\RktCmt{~}\RktCmt{(3) transform the original pattern}

\mbox{\hphantom{\Scribtexttt{xxxxxxxxxxxxx}}}\RktPn{[}\RktSym{new{-}pat}\mbox{\hphantom{\Scribtexttt{x}}}\RktPn{(}\RktSym{transformer}\mbox{\hphantom{\Scribtexttt{x}}}\RktRdr{\#{'}}\RktPn{(}\RktSym{expander}\mbox{\hphantom{\Scribtexttt{x}}}\RktPn{{\hbox{\texttt{.}}} }\RktSym{rest}\RktPn{)}\RktPn{)}\RktPn{]}\RktPn{)}

\mbox{\hphantom{\Scribtexttt{xxxxxxxxx}}}\RktCmt{;}\RktCmt{~}\RktCmt{(4) recur on the new pattern}

\mbox{\hphantom{\Scribtexttt{xxxxxxxxx}}}\RktPn{(}\RktSym{parse{-}pat}\mbox{\hphantom{\Scribtexttt{x}}}\RktSym{new{-}pat}\mbox{\hphantom{\Scribtexttt{x}}}\RktSym{rhs}\RktPn{)}\RktPn{)}

\mbox{\hphantom{\Scribtexttt{xxxxxxx}}}\RktPn{(}\RktSym{\RktValLink{error}}\mbox{\hphantom{\Scribtexttt{x}}}\RktVal{"not an expander"}\RktPn{)}\RktPn{)}\RktPn{)}\RktPn{]}\end{SingleColumn}\end{RktBlk}\end{SCodeFlow}

There are four parts to this new clause.  In step 1, we look up the value
bound to the name \RktSym{expander}.  If this is a \RktSym{match{-}expander} value,
then we extract the transformer function in step 2, which we defined to be a
function from the syntax of a pattern to syntax for a new pattern.  In
step 3, we apply this transformer to the original pattern, producing a
new pattern.  Finally, in step 4, we recur with the \RktSym{parse{-}pat} function, since
we now have a new pattern to be handled.

This loop executes the same steps as the basic loop of macro expansion
as presented by Dybvig et al.\relax{~\cite[Figure 4]{dhb:sc}} and by
Culpepper and Felleisen\relax{~\cite[Figure 2]{cf:scp}}.  This is
unsurprising, since we have extended the syntax of an embedded
language, that of \RktSym{\RktStxLink{match}} patterns, with transformer functions, just as
macros extend the syntax of the expression language with
transformers.  Based on this insight, we can use the same strategies
to ensure \textit{hygiene}, a key correctness criterion for macro
systems, that are applied in existing systems.  Racket provides an API
to these facilities, but their use is beyond the scope of this paper.

These 8 lines of code are the key to extensible pattern matching.
They provide a simple facility for defining arbitrary syntactic
transformations on patterns.  In combination with expressive patterns
such as \RktSym{\badlink{\RktValLink{app}}}, \RktSym{\RktStxLink{and}}, and \RktKw{{\hbox{\texttt{?}}}}, new pattern abstractions and new
uses of pattern matching can be created.

\subsection[Using \RktSym{\RktStxLink{match}} expanders]{Using \RktSym{\RktStxLink{match}} expanders}\label{t:x28part_x22Usingx5fmatchx5fexpandersx22x29}

\begin{Centerfigure}\begin{FigureInside}\begin{SCodeFlow}\begin{RktBlk}\begin{SingleColumn}\RktPn{(}\RktSym{\RktStxLink{define{-}match{-}expander}}\mbox{\hphantom{\Scribtexttt{x}}}\RktSym{\badlink{\RktValLink{polar}}}

\mbox{\hphantom{\Scribtexttt{xx}}}\RktPn{(}\RktSym{\RktStxLink{$\lambda$}}\mbox{\hphantom{\Scribtexttt{x}}}\RktPn{(}\RktSym{stx}\RktPn{)}

\mbox{\hphantom{\Scribtexttt{xxxx}}}\RktPn{(}\RktSym{\RktStxLink{syntax{-}parse}}\mbox{\hphantom{\Scribtexttt{x}}}\RktSym{stx}

\mbox{\hphantom{\Scribtexttt{xxxxxx}}}\RktPn{[}\RktPn{(}\RktSym{\RktStxLink{{\char`\_}}}\mbox{\hphantom{\Scribtexttt{x}}}\RktSym{r}\mbox{\hphantom{\Scribtexttt{x}}}\RktSym{pats}\mbox{\hphantom{\Scribtexttt{x}}}\RktSym{\RktStxLink{{\hbox{\texttt{.}}}{\hbox{\texttt{.}}}{\hbox{\texttt{.}}}}}\RktPn{)}

\mbox{\hphantom{\Scribtexttt{xxxxxxx}}}\RktRdr{\#{'}}\RktPn{(}\RktSym{\RktStxLink{or}}\mbox{\hphantom{\Scribtexttt{x}}}\RktPn{(}\RktSym{\RktValLink{list}}\mbox{\hphantom{\Scribtexttt{x}}}\RktVal{{'}}\RktVal{polar}\mbox{\hphantom{\Scribtexttt{x}}}\RktSym{r}\mbox{\hphantom{\Scribtexttt{x}}}\RktPn{(}\RktKw{{\hbox{\texttt{?}}}}\mbox{\hphantom{\Scribtexttt{x}}}\RktSym{\RktValLink{real{\hbox{\texttt{?}}}}}\mbox{\hphantom{\Scribtexttt{x}}}\RktSym{pats}\RktPn{)}\mbox{\hphantom{\Scribtexttt{x}}}\RktSym{\RktStxLink{{\hbox{\texttt{.}}}{\hbox{\texttt{.}}}{\hbox{\texttt{.}}}}}\RktPn{)}

\mbox{\hphantom{\Scribtexttt{xxxxxxxxxxxxx}}}\RktPn{(}\RktSym{\RktValLink{cons}}\mbox{\hphantom{\Scribtexttt{x}}}\RktVal{{'}}\RktVal{cart}

\mbox{\hphantom{\Scribtexttt{xxxxxxxxxxxxxxxxxxx}}}\RktPn{(}\RktSym{\badlink{\RktValLink{app}}}\mbox{\hphantom{\Scribtexttt{x}}}\RktPn{(}\RktSym{\RktStxLink{$\lambda$}}\mbox{\hphantom{\Scribtexttt{x}}}\RktPn{(}\RktSym{x}\RktPn{)}

\mbox{\hphantom{\Scribtexttt{xxxxxxxxxxxxxxxxxxxxxxxxxxx}}}\RktPn{(}\RktSym{\RktValLink{cons}}\mbox{\hphantom{\Scribtexttt{x}}}\RktPn{(}\RktSym{\RktValLink{sqrt}}\mbox{\hphantom{\Scribtexttt{x}}}\RktPn{(}\RktSym{\RktValLink{apply}}\mbox{\hphantom{\Scribtexttt{x}}}\RktSym{\RktValLink{+}}\mbox{\hphantom{\Scribtexttt{x}}}\RktPn{(}\RktSym{\RktValLink{map}}\mbox{\hphantom{\Scribtexttt{x}}}\RktSym{\RktValLink{sqr}}\mbox{\hphantom{\Scribtexttt{x}}}\RktSym{x}\RktPn{)}\RktPn{)}\RktPn{)}

\mbox{\hphantom{\Scribtexttt{xxxxxxxxxxxxxxxxxxxxxxxxxxxxxxxxx}}}... compute angles ...\RktPn{)}\RktPn{)}

\mbox{\hphantom{\Scribtexttt{xxxxxxxxxxxxxxxxxxxxxxxx}}}\RktPn{(}\RktSym{\RktValLink{list}}\mbox{\hphantom{\Scribtexttt{x}}}\RktSym{r}\mbox{\hphantom{\Scribtexttt{x}}}\RktPn{(}\RktKw{{\hbox{\texttt{?}}}}\mbox{\hphantom{\Scribtexttt{x}}}\RktSym{\RktValLink{real{\hbox{\texttt{?}}}}}\mbox{\hphantom{\Scribtexttt{x}}}\RktSym{pats}\RktPn{)}\mbox{\hphantom{\Scribtexttt{x}}}\RktSym{\RktStxLink{{\hbox{\texttt{.}}}{\hbox{\texttt{.}}}{\hbox{\texttt{.}}}}}\RktPn{)}\RktPn{)}\RktPn{)}\RktPn{)}\RktPn{]}\RktPn{)}\RktPn{)}\RktPn{)}\end{SingleColumn}\end{RktBlk}\end{SCodeFlow}

\Legend{\label{t:x28counter_x28x22figurex22_x22figx3apolarx22x29x29}Figure~3: A match expander}\end{FigureInside}\end{Centerfigure}

In combination with the \RktSym{\badlink{\RktValLink{app}}} pattern, we can now define the
\RktSym{\badlink{\RktValLink{polar}}} example from the introduction.  The implementation is given
in Figure~3.
This expander makes use of the \RktSym{\RktStxLink{or}} pattern to handle either
polar or Cartesian coordinates, and the \RktSym{\badlink{\RktValLink{app}}} pattern to
transform the Cartesian coordinates into polar ones.

\RktSym{\RktStxLink{match}} expanders have been put to numerous other uses in Racket.  For
example, they are used to create a domain{-}specific matching language
for specifying rewrite rules in the PLT web server~\cite{plt-web}.
They are used to provide abstract accessors to complicated data
structures in numerous systems.  A library for defining Wadler{-}style
"views" using \RktSym{\RktStxLink{match}} expanders is available for Racket as well.

\subsection[Further Extensions]{Further Extensions}\label{t:x28part_x22extensionsx22x29}

Based on this framework, we can add additional features to \RktSym{\RktStxLink{match}} expanders,
making them more useful in real{-}world applications.  Above, we discuss
the ability to specify a separate transformer for legacy
pattern{-}matching syntax, allowing the same \RktSym{\RktStxLink{match}} expander to be used with
both syntaxes.  Additionally, since Racket supports creating
structures which can be used as procedures\relax{\cite[Section
3.17]{plttr1}}, \RktSym{\RktStxLink{define{-}match{-}expander}} supports creating \RktSym{\RktStxLink{match}}
expanders which can be also used in expressions, allowing the same
name to be both a constructor and a pattern matching form.

\sectionNewpage

\section[History and Related Work]{History and Related Work}\label{t:x28part_x22relatedx22x29}

\subsection[A History of Pattern Matching in Scheme]{A History of Pattern Matching in Scheme}\label{t:x28part_x22Ax5fHistoryx5fofx5fPatternx5fMatchingx5finx5fSchemex22x29}

To the best of the author{'}s knowledge, the first pattern matcher in
Scheme was written by Matthias Felleisen in February or March of 1984,
inspired by the pattern matching features of Prolog~\cite{prolog-history}.
This original system was written in, or ported to, Kohlbecker et al{'}s
\RktSym{extend{-}syntax} macro system~\cite{kffd:hygiene}, and went by the
name of \RktSym{\RktStxLink{match}}, which was also the name of the primary macro.
The system was then maintained by Bruce Duba until 1991, when Andrew
Wright began maintaining it.  At this point, it was ported to a number
of other macro systems, including Common Lisp{-}style \RktSym{defmacro}~\cite{cl},
Clinger and Rees{'}s explicit renaming system~\cite{wc:macro-renaming}, and
Dybvig et al{'}s expander{-}passing style~\cite{dfh:eps}.  In 1995,
Wright and Duba
prepared an unpublished
manuscript~\cite{wright-tr}\footnote{Although this is often cited as
a Rice University technical report, it was never published as such.} describing the
features of \RktSym{\RktStxLink{match}}, including the grammar of patterns. This
matcher generated decision trees rather than automata.

At this point, the development of the \RktSym{\RktStxLink{match}} library, usually
known as the {``}Wright matcher{''}, stagnated.  Wright{'}s implementation
was ported to \RktSym{\RktStxLink{syntax{-}case}}~\cite{dhb:sc} by Bruce Hauman in
Racket, but this implementation was not used by any other Scheme
implementation.  Hauman also added a number of features not found in
other versions of Wright{'}s library.  Hauman{'}s implementation was then
maintained by the author, and extended with \RktSym{\RktStxLink{match}} expanders
as described in \ChapRefUC{4}{Extensible Patterns}. The author subsequently created
a new implementation in 2008 using backtracking automata, which is
currently distributed with Racket.

In 1990, Quinnec~\cite{cq:match} presented a pattern matcher for
S{-}expressions as well as a formal semantics for his matcher.  However,
we know of no Scheme implementation which distributes his
implementation.

Eisenberg, Clinger and Hartheimer also included a pattern matcher in
their introductory book, \textit{Programming in
MacScheme}~\cite{macscheme-book}. This matcher was implemented as a
procedure rather than a macro, functioned on quoted lists as patterns,
and did not bind variables.

Of course, Scheme macro systems, beginning with Kohlbecker and Wand{'}s
\textit{Macro by Example}~\cite{kw:mbe} have included sophisticated
pattern matchers, functioning entirely on S{-}expressions.  These
systems introduced the \RktSym{\RktStxLink{{\hbox{\texttt{.}}}{\hbox{\texttt{.}}}{\hbox{\texttt{.}}}}} notation for sequence patterns,
later adopted by Wright{'}s matcher among others and seen in the second
example in the introduction.  However, these have
typically not been integrated into the rest of the programming
language and have not made use of sophisticated pattern compilation
techniques.  An exception is Culpepper et al.{'}s \RktSym{\RktStxLink{syntax{-}parse}}
system~\cite{cf:syntax-parse}, which uses a novel compilation
strategy to support sophisticated error reporting.

Finally, numerous Scheme and Lisp systems have implemented their own simple
pattern matching libraries, too many to list here.

\subsection[Other Extensible Pattern Matchers]{Other Extensible Pattern Matchers}\label{t:x28part_x22Otherx5fExtensiblex5fPatternx5fMatchersx22x29}

Numerous other proposals for extensible pattern matching in functional
languages have been presented.  Here, we survey only the most significant.

The original proposal for extensible pattern matching is Wadler{'}s
"views"~\cite{views}, presented in the context of Haskell.  In this
system, a view is a pair of an injection and projection from an
abstract type to an algebraic datatype.  Views are intended to support
data abstraction in conjunction with pattern matching.

Expressing views using \RktSym{\RktStxLink{match}} expanders is straightforward, as is seen
with the example of \RktSym{\badlink{\RktValLink{cart}}} and \RktSym{\badlink{\RktValLink{polar}}}, which form a simple view
on tagged lists.  More complex examples are also expressible,
including the use of views as injections, which uses the
extensions discussed in \SecRefUC{4.4}{Further Extensions}.  Cobbe~\cite{c:views}
provides a library which implements view definitions as a layer on
top of \RktSym{\RktStxLink{match}} expanders.

Also in Haskell, Peyton Jones presents view
patterns~\cite{spj:view-patterns} as an extension to the GHC compiler
and gives a wide range of motivating examples.  These view patterns
are almost identical to \RktSym{\badlink{\RktValLink{app}}} patterns in \RktSym{\RktStxLink{match}}, with the exception
that Peyton Jones suggests using typeclasses to implicitly provide the
function when it is not supplied.  However, this extension is not
implemented in GHC.  Peyton Jones also lists 5 desirable properties of
pattern matching extensions, all of which are provided by \RktSym{\RktStxLink{match}} with
\RktSym{\badlink{\RktValLink{app}}} patterns and \RktSym{\RktStxLink{match}} expanders.

Active patterns, originally proposed by
Erwig~\cite{erwig:active-patterns} and subsequently extended and
implemented in F\# by Syme et al.~\cite{snm:active-patterns}, as well
as Scala extractors~\cite{scala-extractors}, provide
more expressive extensions.  Users can define both partial and total
patterns.  Partial patterns can be implemented as an abstraction over
\RktSym{\badlink{\RktValLink{app}}} and \RktKw{{\hbox{\texttt{?}}}} patterns, using a helper function.  Total patterns
require the definition of several \RktSym{\RktStxLink{match}} expanders, each using such an
abstraction, with only one helper function.  Using both pattern
abstraction and abstraction over definitions provided by macros, such extensions
can be specified in Racket just as in F\#, but again without the need
to modify the base language.   Since \RktSym{\RktStxLink{match}} does not check
exhaustiveness of pattern matching, total active patterns cannot be
verified to match completely.

\sectionNewpage

\section[Conclusion]{Conclusion}\label{t:x28part_x22Conclusionx22x29}

Pattern matching is an invaluable tool for programmers to write
concise, maintainable, and performant code.  However, it does not
usually support the abstraction facilities that functional programmers
expect in other parts of the language.  In this paper, we describe a
syntactic abstraction facility that allows arbitrary rewriting of
patterns at compilation time.  Furthermore, this is provided in a
pattern matching system that is implemented as a syntactic extension
in Racket.  The implementation reveals a striking similarity between
the base language extension mechanism and extensions defined in
higher{-}level domain{-}specific languages such as pattern matching.

\relax{\subsection*{Acknowledgments}}

Ryan Culpepper provided invaluable assistance in the development of
\RktSym{\RktStxLink{match}}, and devised the algorithm used for handling sequence patterns.  Matthias
Felleisen shared his knowledge of the early history of pattern
matching in Scheme. Stevie Strickland and Vincent St{-}Amour provided
feedback on earlier drafts of the paper.  The author is supported by a
grant from the Mozilla Foundation.

\bibliographystyle{plain}\bibliography{new-popl}

\postDoc
\end{document}